\documentclass[%
 pra,reprint,
 amsmath,amssymb,
 aps,
floatfix,superscriptaddress
]{revtex4-2}
\usepackage{float} 
\usepackage{graphicx}
\usepackage{dcolumn}
\usepackage{bm}
\usepackage{siunitx}
\usepackage{physics}
\usepackage{xcolor}
\usepackage[caption=false]{subfig}
\usepackage[export]{adjustbox}

\usepackage{hyperref}
\hypersetup{
        colorlinks=false
}

\usepackage{nicematrix}
\NiceMatrixOptions{
code-for-first-row = \color{black} ,
code-for-last-row = \color{black} ,
code-for-first-col = \color{black} ,
code-for-last-col = \color{black}
}

\AtBeginDocument{\RenewCommandCopy\qty\SI} 
\newcommand{\bop}{\hat{b}}
\newcommand{\bdag}{\hat{b}^\dagger}

\newcommand{\aop}{\hat{a}}
\newcommand{\adag}{\hat{a}^\dagger}
\newcommand{\adagT}{\hat{a}^{\dagger 2}}
\newcommand{\adagTh}{\hat{a}^{\dagger 3}}
\newcommand{\adagn}{\hat{a}^{\dagger n}}
\newcommand{\sigP}{\hat{\sigma}_+}
\newcommand{\sigM}{\hat{\sigma}_-}
\newcommand{\sigZ}{\hat{\sigma}_z}
\newcommand{\sigX}{\hat{\sigma}_x}
\newcommand{\sigY}{\hat{\sigma}_y}

\begin{document}

\preprint{APS/123-QED}

\title{Driven Multiphoton Qubit-Resonator Interactions}

\author{Mohammad Ayyash}
\thanks{mmayyash@uwaterloo.ca}

\affiliation{%
 Institute for Quantum Computing, University of Waterloo, 200 University Avenue West, Waterloo,
Ontario N2L 3G1, Canada}
\affiliation{Department of Physics and Astronomy, University of Waterloo, 200 University Avenue West, Waterloo,
Ontario N2L 3G1, Canada}
\affiliation{Red Blue Quantum Inc., 72 Ellis Crescent North, Waterloo, Ontario N2J 3N8, Canada}

\author{Xicheng Xu}
\affiliation{%
 Institute for Quantum Computing, University of Waterloo, 200 University Avenue West, Waterloo,
Ontario N2L 3G1, Canada}
\affiliation{Department of Physics and Astronomy, University of Waterloo, 200 University Avenue West, Waterloo,
Ontario N2L 3G1, Canada}
\affiliation{Red Blue Quantum Inc., 72 Ellis Crescent North, Waterloo, Ontario N2J 3N8, Canada}

\author{Sahel Ashhab}

\affiliation{Advanced ICT Research Institute, National Institute of Information and
Communications Technology, 4-2-1, Nukui-Kitamachi, Koganei, Tokyo 184-8795, Japan}

\author{M. Mariantoni}
\affiliation{%
 Institute for Quantum Computing, University of Waterloo, 200 University Avenue West, Waterloo,
Ontario N2L 3G1, Canada}
\affiliation{Department of Physics and Astronomy, University of Waterloo, 200 University Avenue West, Waterloo,
Ontario N2L 3G1, Canada}
\affiliation{Red Blue Quantum Inc., 72 Ellis Crescent North, Waterloo, Ontario N2J 3N8, Canada}

\date{\today}
    
\begin{abstract}
   We develop a general theory for multiphoton qubit-resonator interactions enhanced by a qubit drive. The interactions generate qubit-conditional operations in the resonator when the driving is near $n$-photon cross-resonance, i.e., when the qubit drive is $n$-times the resonator frequency. We pay special attention to the strong driving regime, where the resulting effective interactions are conditioned on the qubit dressed states. Next, we investigate the use of a two-tone drive to engineer an effective $n$-photon Rabi Hamiltonian {with widely tunable effective system parameters, which could enable} the realization of new regimes that have so far been inaccessible. Then, we discuss applications for the specific case where $n=2$, which results in qubit-conditional squeezing (QCS). We show that the QCS protocol {can be used to generate} a superposition of orthogonally squeezed states following a properly chosen qubit measurement. We outline quantum information processing applications for these states, including encoding a qubit in a resonator via the superposition of orthogonally squeezed states. {We show how the QCS operation can be used to realize a controlled-squeeze gate and its use in bosonic phase estimation. The QCS protocol can also be utilized to achieve faster unitary operator synthesis on the joint qubit-resonator Hilbert space.} Finally, we propose a multiphoton circuit QED implementation based on a transmon qubit coupled to a resonator via an asymmetric SQUID. We provide realistic parameter estimates for the two-photon operation regime that can host the aforementioned two-photon protocols. We use numerical simulations to show that even in the presence of spurious terms and decoherence, our analytical predictions are robust.
\end{abstract}

\maketitle



\section{Introduction}
In the last century, mastering the manipulation of quantum-mechanical light-matter interactions emerged as a groundbreaking achievement. Today, the focus has evolved towards the precise engineering of versatile interactions, resilient to decoherence and practical imperfections, essential for advancing quantum technologies. This pursuit has the potential to advance information processing and error correction, and ultimately, it could lead to the realization of fault-tolerant quantum computing. Moreover, the precise control of light-matter interactions extends far beyond computing, finding diverse applications in quantum metrology, communication, and simulations, highlighting its profound impact across various domains.

The elementary model of quantum light-matter interactions is captured by the Rabi model where a qubit is linearly coupled to a single quantized field mode or a resonator \cite{JaynesCummings,walls1994gj,Sahel_QO_Nonclassical,RabiOverview_Braak}. This model describes the basic physics underlying most quantum computing implementations. This includes circuit quantum electrodynamics (QED) \cite{CircuitQEDReview}, trapped ions \cite{TrappedIonsReview}, and cavity QED \cite{Haroche_Raimond}.

Different variants of the Rabi model exhibit a variety of higher order perturbative multiphoton effects stemming from a linear interaction (see for example Refs.~\cite{Multiphoton0,Multiphoton1,Multiphoton2,Multiphoton3,Multiphoton4,Multiphoton6,MM_TwoPhotonJC}).  These multiphoton perturbative effects have proven their utility in various applications, e.g. improved readout due to qubit-induced nonlinearity \cite{NonlinearReadout}. Thus, to further control and leverage multiphoton effects, the Rabi model can be generalized to include nonlinear interactions, namely, a qubit nonlinearly coupled to a resonator through an $n$-photon interaction. These nonlinear models are nonperturbative, as the nonlinear interaction is inherent to the Hamiltonian rather than higher-order effects of a linear interaction term. Some of the spectral and dynamical properties of multiphoton Rabi models describing such nonlinear interactions, e.g. two-photon interactions, have been previously studied \cite{QStat_NonlinearOptics,DFWalls_1971,CEmary_2002,GenRabi2,GenRabi3,GenRabi4,GenRabi5,GenRabi6,GenRabi7,GenRabi8,GenRabi9}. Other studies of these models were focused on multiphoton blockades \cite{ImplementationSC1,MultiphotonJC_PRL,ImplementationSC3}, `Fock state filters' that effectively confine the dynamics to a finite-dimensional subspace \cite{MultiphotonJC_PRL}, enhancement of collective multiqubit phenomena \cite{TwoPhotonBadCavity} and stabilization of nonclassical states for quantum error correction \cite{CombinedConf}. Towards the goal of experimental realization, a series of nonperturbative implementations of the two-photon Rabi model have been recently proposed in superconducting circuits \cite{ImplementationSC1,ImplementationSC2,ImplementationSC3} and trapped ions \cite{ImplementationTI1,ImplementationTI2}.

Much remains to be discovered about the various regimes of nonperturbative multiphoton qubit-resonator interactions, particularly when the qubit or resonator is driven, since the driving alters these interactions. In this paper, we develop a general theory for driving-enhanced nonperturbative multiphoton interactions in a qubit-resonator system. In particular, we study a qubit nonlinearly coupled to a resonator through an $n$-photon Rabi interaction in the presence of a qubit drive. 

The paper is structured as follows: Sec.~\ref{sec:DrivenInt} lays out the formalism for the theory. Then, the driving regimes on- and off-resonance from the qubit and resonator are explored. The driving is found to generate qubit-conditional operations on the resonator. {Next, we use two-tone driving to engineer an effective $n$-photon Rabi model that is tunable to arbitrary coupling strengths, thereby performing a quantum simulation of the model.} In Sec.~\ref{sec:TwoPhApp}, we apply the developed theory to the case of $n=2$, where we discover a \textit{qubit-conditional squeezing} (QCS) process. The QCS protocol allows for the encoding of a qubit state in the superposition of orthogonally squeezed states in the resonator. {Then, we show the potential use of the QCS protocol in the phase estimation algorithm with bosonic systems. We describe how the generators of the QCS operation can be used to realize faster unitary synthesis on the joint qubit-resonator Hilbert space.} We discuss how these applications can be generalized for higher-order interactions. Section~\ref{sec:cQED} proposes an implementation scheme based on the transmon qubit which can host the required two-photon interaction for implementing the QCS protocols. Lastly, we summarize our findings and present an outlook in Sec.~\ref{sec:Conc}.

\section{Enhancing multiphoton interactions with cross-resonant driving}\label{sec:DrivenInt}

In this section, we develop the theory of driving-enhanced interactions that enables qubit-conditional resonator operations. We proceed by stating the system and drive Hamiltonians and applying the necessary transformations to simplify its time dependence. Once we arrive at a simplified effective Hamiltonian, using the dressed basis, we explore the dynamics and its implications. Then, we add a second drive to the qubit, with properly chosen values of the amplitude and frequency, to engineer an effective $n$-photon Rabi Hamiltoninian with tunable parameters allowing the access to arbitrary coupling regimes.

\subsection{System Hamiltonian}
We start by considering the driven $n$-photon Rabi model whose Hamiltonian reads
\begin{subequations}\label{eq:SysHam}
\begin{align}
    \hat{H}=\hat{H}_{n-\text{R}} + \hat{H}_d
\end{align}
where
\begin{align}
    \hat{H}_{n-R}=\frac{\hbar \omega_q}{2}\hat{\sigma}_z+ \hbar\omega_r \hat{a}^\dagger\hat{a} + \hbar g_n(\hat{\sigma}_+ + \hat{\sigma}_-)(\hat{a}^{\dagger n} + \hat{a}^n),
\end{align}
and
\begin{align}
    \hat{H}_d=\hbar\Omega \cos(\omega_d t)(\sigP + \sigM).
\end{align}
\end{subequations}
Here, $\hat{\sigma}_z=\dyad{e}-\dyad{g}$ describes the population difference between the excited energy state $\ket{e}$ and the ground state $\ket{g}$ of the qubit, $\hat{\sigma}_+=\dyad{e}{g}$ and $\hat{\sigma}_-=\hat{\sigma}_+^{\dagger}$ are raising and lowering operators of the qubit, $\hat{a}$ and $\hat{a}^\dagger$ are the annihilation and creation operators of the resonator, $\omega_q$ is the transition frequency of the qubit, $\omega_r$ is the resonance frequency of the resonator, $g_n$  is the $n$-photon coupling strength between the resonator and qubit, $\Omega$ is the strength of the driving field and $\omega_{\text{d}}$ is the driving frequency. 

We rewrite the Hamiltonian of Eq.~\eqref{eq:SysHam} in a particular rotating frame, accounting for the $n$-photon nature of the qubit-resonator interaction, by means of the unitary transformation $\hat{U}^{r,n}=\exp[-i\omega_{\text{d}}t(\hat{\sigma}_z/2 + \hat{a}^\dagger \hat{a}/n)]$,
\begin{align}\label{eq:FullRotFrameHam}
    \hat{H}^r =& \hat{U}^{r,n \dagger} \hat{H}\hat{U}^{r,n}+i\hbar \dot{\hat{U}}^{r,n \dagger}\hat{U}^{r,n}\nonumber\\ =&\frac{\hbar \Delta}{2}\hat{\sigma}_z + \hbar \delta_n\hat{a}^\dagger \hat{a} \nonumber\\&+ \hbar g_n\big(\hat{\sigma}_+\hat{a}^n + \hat{\sigma}_-\hat{a}^{\dagger n}\nonumber\\ &\,\,\,\,\,\,\,\,\,\,\,\,\,\,\,\,\,\,\,\,+ e^{i2\omega_d t}\sigP\adagn + e^{-i2\omega_d t} \sigM\aop\big)\nonumber\\& +\frac{\hbar\Omega}{2}(\sigP +\sigM +e^{i2\omega_d t}\sigP + e^{-i2\omega_d t}\sigM), 
\end{align}
where $\Delta=\omega_q -\omega_d$ and $\delta_n=\omega_r - \omega_d/n$. We may now simplify this Hamiltonian by imposing the rotating-wave approximation (RWA) condition,
    \begin{align}
        &\label{eq:RWA2}g_n, \,\Omega,\, \Delta,\,\delta_n \ll \omega_d.
    \end{align}
This condition is neccesary to eliminate the fast-oscillating counter-rotating interaction terms, $g_n(e^{+i2\omega_{\text{d}}t}\hat{\sigma}_+\adagn +e^{-i2\omega_{\text{d}}t}\hat{\sigma}_-\hat{a}^n)$, and counter-rotating driving terms, $\Omega(e^{+i2\omega_{\text{d}}t}\hat{\sigma}_+ +e^{-i2\omega_{\text{d}}t}\hat{\sigma}_-)/2$. Imposing these RWA conditions, the simplified Hamiltonian reads

\begin{align}\label{eq:SimplifiedRotFrameHam}
    \hat{H}^r_{\text{RWA}} =&\frac{\hbar \Delta}{2}\hat{\sigma}_z +\frac{\hbar \Omega}{2}\hat{\sigma}_x + \hbar \delta_n\hat{a}^\dagger \hat{a} \nonumber\\&+ \hbar g_n(\hat{\sigma}_+\hat{a}^n +\hat{\sigma}_-\hat{a}^{\dagger n}), 
\end{align}
where $\hat{\sigma}_x=\hat{\sigma}_++\hat{\sigma}_-$. This last Hamiltonian will serve as the basis for our study.

\subsection{Effective interaction\label{sec:EffectiveInt}}
The qubit-resonator interaction changes depending on the driving parameters. We now aim to investigate the dynamics within two driving regimes, focusing on how the drive affects the qubit-resonator interaction. To better understand the driving regime's effect on the interaction and further simplify the analytical calculations, we transform to the interaction picture using the unitary $\hat{U}^{(I)}=\exp[-i \hat{H}_0 t/\hbar]$, where $\hat{H}_0=\hbar\Delta \hat{\sigma}_z/2 + \hbar\Omega\hat{\sigma}_x/2 + \hbar \delta_n \hat{a}^\dagger \hat{a}$. The interaction picture Hamiltonian reads
\begin{align}\label{eq:IntPicHam}
     \hat{H}^{(I)} =& \hat{U}^{(I) \dagger} \hat{H}^r_{\text{RWA}}\hat{U}^{(I)}+i\hbar \dot{\hat{U}}^{(I) \dagger}\hat{U}^{(I)}\nonumber\\ =&\hbar g_n\bigg[ \frac{\sin(\theta)}{2}\left( \dyad{\overline{+}}-\dyad{\overline{-}}\right) \nonumber\\&+\cos^2\left(\frac{\theta}{2}\right)e^{i\varepsilon t}\dyad{\overline{+}}{\overline{-}}\nonumber\\ &-\sin^2\left(\frac{\theta}{2}\right)e^{-i\varepsilon t}\dyad{\overline{-}}{\overline{+}}\bigg] \hat{a}^n e^{-in\delta_n t} + \text{H.c.},
\end{align}
where we use the dressed states $\ket{\overline{+}}=\sin\left(\theta /2\right)\ket{{g}}+\cos\left(\theta /2\right)\ket{{e}}$ and $\ket{\overline{-}}=\cos\left(\theta /2\right)\ket{{g}}-\sin\left(\theta /2\right)\ket{{e}}$, $\varepsilon=\sqrt{\Omega^2 +\Delta^2}$ and $\theta=\arctan(\Omega/\Delta)$.

The Hamiltonian of Eq.~\eqref{eq:IntPicHam} reveals two distinct interactions taking place at different timescales. One of these interactions oscillates with $e^{\pm i \varepsilon t}$; as the driving strength, $\Omega$, increases, these terms oscillate rapidly. We can eliminate these fast-oscillating terms by imposing the driving-detuning RWA condition \begin{align}\label{eq:DrivCond}
    |n\delta_n|,g_n\ll \varepsilon.
\end{align}
Imposing this condition allows us to obtain the effective Hamiltonian
\begin{align}\label{eq:EffHam}
\hat{H}^{(I)}_{\text{eff}}=\hbar \overline{g}_n (\dyad{\overline{+}}-\dyad{\overline{-}})(\hat{a}^{\dagger n} e^{ni\delta_n t}+\hat{a}^n e^{-ni\delta_n t}),
\end{align}
where $\overline{g}_n=g_n\sin(\theta)/2.$
The dynamics associated with this Hamiltonian result in a conditional $n$-photon {resonator operation dependent on the qubit state}. { The time-evolution operator generated by Eq.~\eqref{eq:EffHam} is \begin{align}\label{eq:GenTimeEvOp}
\hat{U}_{\text{eff},n}(t,0)=\dyad{\overline{+}}\hat{S}_n(\lambda(t)) + \dyad{\overline{-}}\hat{S}_n(-\lambda(t)), 
\end{align}
where $\hat{S}_n(\lambda)=\exp((\lambda^* \aop^n - 
\lambda \adagn)/n! )$ is the generalized $n$-photon squeezing operator \cite{GeneralizedSqueezing}; for $n=1$ it is the usual displacement operator, for $n=2$ it is the squeezing operator, etc., and $\lambda(t)=\overline{g}_n n! (e^{in \delta_n t}-1)/2n\delta_n $ is the generalized $n$-photon squeezing parameter.}

The driving-detuning condition can be achieved by changing $\Omega$ and $\Delta$ such that Eq.~\eqref{eq:DrivCond} is satisfied. The dressed basis states also depend on $\Omega$ and $\Delta$, and depending on the parameter regime, they can be approximated as the $\sigX$ or $\sigZ$ bases. In what follows, we explore the two extremes of strong driving and qubit-detuned weak driving. 
\subsubsection{Strong driving regime}\label{sec:StrongDriv}
When the driving is strong, $\Omega \gg \Delta$, $\ket{\overline{\pm}}\simeq\ket{\pm}=(\ket{g}\pm\ket{e})/\sqrt{2}$, i.e., the dressed basis is the $\sigX$ basis.
 In this case, the effective Hamiltonian is \begin{align}\label{eq:StrongDriv}
\hat{H}_{\text{eff}}^{(I)}\simeq\hbar\overline{g}_n\sigX(\hat{a}^{\dagger n} e^{ni\delta_n t}+\hat{a}^n e^{-ni\delta_n t}).
\end{align}
Note that this last equation becomes exact when $\Delta=0$, since in this case, $\varepsilon=\Omega$ and $\sin(\theta/2)=\cos(\theta/2)=1/\sqrt{2}$. In this strong driving regime, the multiphoton interaction is conditioned on the basis $\{\ket{+},\ket{-}\}$.

The Hamiltonian of Eq.~\eqref{eq:StrongDriv} admits another useful interpretation, namely, the strong driving effectively places the co-rotating ($n$-photon JC) terms, $\sigP\aop^n+\sigM\adagn$, and the counter-rotating ($n$-photon anti-JC) terms, $\sigP\adagn +\sigM\aop^n$, being on the same timescale. In general, when the co-rotating and counter-rotating interactions are on the same timescale, we get effective interactions that generate qubit-conditional operations.

The case of $n=1$ yields qubit-conditional displacements of the resonator state \cite{ResSchCats,SolanoCat}. This is similar to other dispersive techniques in which the resonator is strongly driven, leading to qubit-conditional displacements \cite{QECGrid2020,FastUnivControlDisp}.
When $n=2$, Eq.~\eqref{eq:GenTimeEvOp} performs qubit-conditional squeezing, which will be the primary focus of Sec.~\ref{sec:TwoPhApp}. For $n=3$, the effective interactions result in qubit-conditional `trisqueezing'. Unconditional trisqueezing has been recently achieved in superconducting circuits \cite{CWilson2,Trisq2}. Additionally, unconditional triqsqueezing and quadsqueezing ($n=4$) have been realized in a trapped ions implementation \cite{UncondSqTrappedIons}. Trisqueezed states can be used to generate resource states for continuous-variable universal quantum computation \cite{CubicPhase}. In general, resonator states generated by $n$-photon interactions (for $n>2$) acting on the vacuum are typically used as non-Gaussian resource states for quantum computation.

\subsubsection{Qubit-detuned weak driving regime}

The driving-detuning RWA performed on Eq.~\eqref{eq:IntPicHam} to obtain Eq.~\eqref{eq:EffHam} relies on the condition $\varepsilon=\sqrt{\Omega^2+ \Delta^2}\gg |n\delta_n|,g_n$, which can be satisfied even for weak driving with a large qubit detuning that keeps $\varepsilon$ large. When $|\Delta|\gg \Omega$, $\ket{\overline{+}}\simeq\ket{e}$ and $\ket{\overline{-}}\simeq\ket{g}$, and the Hamiltonian of Eq.~\eqref{eq:EffHam} becomes
\begin{align}\label{eq:DetEffHam}
\hat{H}^{(I)}_{\text{eff}}&\simeq\hbar \overline{g}_n (\dyad{e}-\dyad{g})(\hat{a}^{\dagger n} e^{ni\delta_n t}+\hat{a}^n e^{-ni\delta_n t})\nonumber\\ &= \hbar \overline{g}_n \sigZ (\hat{a}^{\dagger n} e^{ni\delta_n t}+\hat{a}^n e^{-ni\delta_n t}),
\end{align}
where the $n$-photon interaction is now conditioned on the qubit state in the bare basis $\{\ket{g},\ket{e}\}.$ In this weak but largely detuned driving regime, the case of $n=1$ where the drive is cross-resonant with the resonator, $\delta_1=0$, corresponds to the well-known cross-resonance readout \cite{GatedConditionalDisp}. Generally, the rate of photon generation in the resonator depends on $\overline{g}_n$, which is greater in the strong driving regime when compared to weak detuned driving.

\subsection{Engineering effective $n$-photon Rabi Hamiltonian with arbitrary coupling strength}\label{sec:QuantumSim}
The $n$-photon Jaynes-Cummings Hamiltonian introduced in Eq.~\eqref{eq:SysHam} is often an excellent approximation, for weak coupling, of the more general $n$-photon Rabi model. The main difference is the presence or absence of the counter-rotating interaction terms, $\propto\sigP\adagn + \sigM\aop^n$. We now use an additional qubit drive on the system to arrive at an effective $n$-photon Rabi model with arbitrary coupling strengths, allowing the access to regimes that are currently unachievable in experimental settings \cite{USCQuantumSim}.

We relabel the drive parameters to distinguish the two drives considered; each drive is characterized by a strength $\Omega_k$ and a driving frequency $\omega_{dk}$ with $k=1,2$. We start by considering the Hamiltonian of Eq.~\eqref{eq:SimplifiedRotFrameHam} in the presence of the two drives, which reads 
\begin{align}\label{eq:SimplifiedRotFrameHam2}
    \hat{\widetilde{H}}^r_{\text{RWA}} =&\frac{\hbar \Delta}{2}\hat{\sigma}_z +\frac{\hbar \Omega_1}{2}\hat{\sigma}_x + \hbar \delta_n\hat{a}^\dagger \hat{a} \nonumber\\&+ \hbar g_n(\hat{\sigma}_+\hat{a}^n +\hat{\sigma}_-\hat{a}^{\dagger n})\nonumber\\ &+\frac{\hbar\Omega_2}{2}(e^{i\delta_d t}\sigP +e^{-i\delta_d t}\sigM), 
\end{align}
where $\Delta=\omega_q-\omega_{d1}$, $\delta_{n}=\omega_r-\omega_{d1}/n$ and $\delta_d=\omega_{d1}-\omega_{d2}$. Here, we note that the Hamiltonian is in the rotating frame with respect to $\omega_{d1}$. Both drives are operated within the RWA regime, where
\begin{align*}
    \Omega_k \ll \omega_{dk}.
\end{align*}
In this setup, we seek to obtain three tunable terms in the effective Hamiltonian; qubit, resonator and interaction terms. The importance of the second drive is that it introduces a qubit term in the final effective Hamiltonian. To elucidate how the second drive plays this role, we make another transformation to the interaction picture with respect to the $\sigX$ term in Eq.~\eqref{eq:SimplifiedRotFrameHam2} via $\hat{U}^{(I)}=\exp[-i \hat{H}_0 t/\hbar]$, where $\hat{H}_0=\hbar\Omega_1\hat{\sigma}_x/2$. This frame is chosen on the basis that we operate in the strong driving regime of the first drive, where $\Omega_1$ is the largest energy scale in Eq.~\eqref{eq:SimplifiedRotFrameHam2}. In this interaction picture, the system Hamiltonian reads
\begin{figure}[t]
        \includegraphics[scale=.8,left,trim={0.17cm .8cm 0 0}]{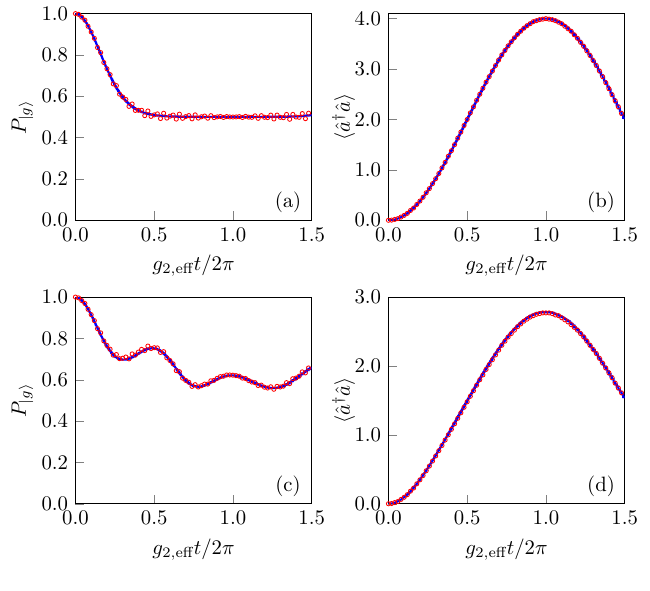}
        \caption{Quantum simulation of the two-photon Rabi model in the ultrastrong coupling regime. The time-evolution of the ground state probability and the resonator photon number are shown for a system initialized in $\ket{g}\ket{0}$. The blue solid line is generated by Eq.~\eqref{eq:TwoDriveEff} and the red circles are generated by Eq.~\eqref{eq:TwoDriveIntPic}. The parameters used are $\Omega_1=\delta_d=2\pi\times\SI{1.4}{\giga\hertz}$, $\Delta=2\pi\times\SI{20}{\mega\hertz}$, $g_{2,\text{eff}}=2\pi\times\SI{10}{\mega\hertz}$, $\omega_{r,\text{eff}}=2\pi\times\SI{10}{\mega\hertz}$. For (a),(b) $\omega_{q,\text{eff}}=0$ and for (c),(d) $\omega_{q,\text{eff}}=2\pi\times\SI{10}{\mega\hertz}$. }
        \label{fig:QuantumSim}
\end{figure}
\begin{align}\label{eq:TwoDriveIntPic}
    \hat{H}^{(I)}=&-\frac{\hbar\Delta}{2}(e^{i\Omega_1 t}\dyad{+}{-}+e^{-i\Omega_1 t}\dyad{-}{+}) +\hbar \delta_{n} \hat{a}^\dagger \hat{a} \nonumber\\ &+\frac{\hbar}{2}\Bigg[ \bigg(\dyad{+}-\dyad{-} +e^{i\Omega_1 t}\dyad{+}{-}\nonumber\\ &-e^{-i\Omega_1 t}\dyad{-}{+}\bigg)\bigg(g_n \hat{a}^n + \frac{\Omega_2}{2}e^{i\delta_d t}\bigg)+\text{H.c.}\Bigg].
\end{align}

As discussed in Sec.~\ref{sec:StrongDriv}, with $\Omega_1$ as the dominant energy scale, we explicitly impose $\Omega_1 \gg |\Delta|, g_n$. This assumption enables us to disregard the rapidly oscillating terms with factors of $e^{\pm i\Omega_1 t}$. Next, we set $\delta_d=\Omega_1$, which cancels the time dependence in the terms responsible for the effective qubit term, $-(\dyad{-}{+} e^{i(\delta_d -\Omega_1) t} + \text{H.c.})$. Note that the term $(\dyad{+}{-} e^{i(\delta_d +\Omega_1)t} + \text{H.c.})$ oscillates with $e^{\pm i (\delta_d + \Omega_1)t}$ and can therefore be ignored. This allows us to obtain an effective $n$-photon Rabi Hamiltonian
\begin{align}\label{eq:TwoDriveEff}
    \hat{H}_{\text{eff}}^{(I)}=&-\frac{\hbar \Omega_2}{4}(\dyad{+}{-}+\dyad{-}{+}) +\hbar \delta_n\hat{a}^\dagger \hat{a} \nonumber\\ &+ \frac{\hbar g_n}{2} (\dyad{+}-\dyad{-})(\hat{a}^{\dagger n} + \hat{a}^n)\nonumber\\ =& \frac{\hbar \omega_{q,\text{eff}}}{2}\hat{\sigma}_z + \hbar \omega_{r,\text{eff}}\hat{a}^\dagger \hat{a} \nonumber\\&+ \hbar g_{n,\text{eff}} \sigX (\hat{a}^{\dagger n} + \hat{a}^n),
\end{align}
where $\omega_{q,\text{eff}}=\Omega_2/2$, $\omega_{r,\text{eff}}=\delta_{n}$ and $g_{n,\text{eff}}=g_n/2$. Here, we have rewritten the Hamiltonian in the bare basis where $\sigX=\sigP+\sigM=\dyad{+}-\dyad{-}$ and $\sigZ=-(\dyad{+}{-}+\dyad{-}{+})/2$.  The effective system parameters are highly tunable and allow for a quantum simulation of the $n$-photon Rabi model in various coupling regimes. Note that when $\Omega_2=0$ (i.e. in the absence of the second drive), Eq.~\eqref{eq:TwoDriveEff} is the same as Eq.~\eqref{eq:StrongDriv} with the only difference being a transformation with respect to the resonator term via $\exp(i\delta_n t \adag \aop)$. Therefore, in the case of $\Omega_2=0$, we recover the results of Sec.~\ref{sec:StrongDriv}.

In the multiphoton generalizations of the Rabi model, {the relationship between the order of the interaction and the critical coupling at which the spectral collapse occurs is unknown}. Thus, an effective Hamiltonian with tunable parameters allows us to probe the dynamical behaviour in such extreme scenarios. Figure~\ref{fig:QuantumSim} shows the dynamics of the effective Hamiltonian in Eq.~\eqref{eq:TwoDriveEff} compared to Eq.~\eqref{eq:TwoDriveIntPic} for the case $n=2$ in the ultrastrong coupling regime { which starts around} $g_{2,\text{eff}}/\omega_{r,\text{eff}}\simeq 0.1$. As the amplitude $\Omega_1$ increases, the effective and full Hamiltonian dynamics get closer to each other. Even for experimentally realistic drive strengths (Fig.~\ref{fig:QuantumSim} uses $\Omega_1=2\pi\times \SI{1.4}{\giga\hertz}$), the dynamics of the full Hamiltonian with the same parameters very closely resembles that of the effective model. 

Increasing the native coupling strength of the system -- as we will see later -- typically comes with an increase in the strength of spurious terms that may completely ruin the desired interaction. Thus, engineering effective Hamiltonians in extreme parameter regimes using appropriately designed driving fields allows for achieving experimentally inaccessible regimes using easily accessible coupling strengths.
\\

\section{Applications\label{sec:TwoPhApp}}

In this section, we focus on the quantum information processing applications stemming from the two-photon interaction generating qubit-conditional squeezing. We also discuss how these applications can be generalized for higher-order interactions where $n>2$.
\\

\subsection{Two-photon interactions}
The effective time-evolution operator of Eq.~\eqref{eq:GenTimeEvOp} results in qubit-state-dependent displacement ($n=1$), squeezing ($n=2$), trisqueezing ($n=3$), etc., whose effects are most pronounced when the driving is ($n$-photon) cross-resonant, $\delta_n=0$, as the relevant parameters grow linearly in time, $i\overline{g}_n t$. The effect of cross-resonance is, therefore, to facilitate the most efficient and sustained channeling of photons from the drive through the qubit into the resonator.

The strong driving regime of the one-photon interaction has been studied in the works of Refs.~\cite{SolanoCat,ResSchCats}. The main outcome, when $n=1$, is the generation of qubit-conditional displacements that allow for the generation of Schr\"odinger cat states. In this section, we explore the case of $n=2$ yielding qubit-conditional squeezing and its applications. Qubit-conditional squeezing has previously been investigated using a different mechanism, which relied on the motional modes of trapped ions to generate the required interaction \cite{SD-MotSqueezing}.

\subsubsection{Schr\"odinger-cat-like superposition of orthogonally squeezed states}
\begin{figure}[t]
        \includegraphics[scale=.8,left,trim={0 0 0 0}]{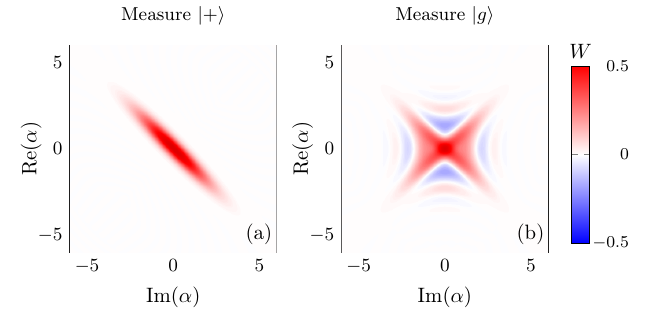}
        \caption{Wigner function heatmap of the resonator state for the case of $n=2$ when measuring the qubit in the dressed vs the bare bases. The resonator state (after a qubit measurement) generated by Eq.~\eqref{eq:QCSTimeOp} after time-evolution period of $g_2t/2\pi=0.15$ for an initial state $\ket{g}\ket{0}$. The parameters used are $g_2=2\pi\times \SI{20}{\mega\hertz}$, and $\Delta=\delta_2=0$. {For simplicity, we set $\Delta=0$ which makes $\overline{g}_2=g_2.$}  (a) The resonator is left in a single well-defined squeezed state when the qubit is measured in the dressed basis. (b) On on the other hand, it is left in a superposition of orthogonally squeezed states when the qubit is measured in the bare basis.}
        \label{fig:WignerFunctions}
\end{figure}

When $n=2$, the time-evolution operator of Eq.~\eqref{eq:GenTimeEvOp} is
\begin{align}\label{eq:QCSTimeOp}
    \hat{U}_{\text{eff}}^{(I)}(t,0)=\dyad{\overline{+}}\hat{S}(\zeta(t))+\dyad{\overline{-}}\hat{S}(-\zeta(t)),
\end{align}
where $\hat{S}(\zeta)=\exp((\zeta^* \hat{a}^2-\zeta \hat{a}^{\dagger 2})/2)$ is the squeezing operator and $\zeta(t)=\overline{g}_2(e^{i2\delta_2 t}-1)/2\delta_2$ is the squeezing parameter; when $\delta_2\rightarrow0$, $\zeta(t)=i\overline{g}_2t$. For simplicity, we henceforth assume $\Delta=0$ such that $\ket{\overline{\pm}}=\ket{\pm}$ \footnote{It is straightforward to use the general dressed basis for what follows. Simply replace $\ket{+}$ with $\ket{\overline{+}}$ and $\ket{-}$ with $\ket{\overline{-}}$ in the prepared states.}. When the system is initialized with the qubit in the ground state and the resonator in vacuum, $\ket{\psi_{\text{i}}}=\ket{g}\ket{0}=(\ket{+}+\ket{-})\ket{0}/\sqrt{2}$, the time-evolved state reads
\begin{align}\label{eq:TimeEvSt}
    \ket{\psi(t)}^{(I)}=&\frac{1}{\sqrt{2}}(\ket{+}\ket{\zeta(t)}+\ket{-}\ket{-\zeta(t)})\nonumber\\ =&\frac{1}{2}\ket{g}(\ket{\zeta(t)}+\ket{-\zeta(t)})\nonumber \\&+\frac{1}{2}\ket{e}(\ket{\zeta(t)}-\ket{-\zeta(t)}),
\end{align}
where $\ket{\zeta}=\hat{S}(\zeta)\ket{0}$ is a squeezed vacuum state.
If the qubit is measured in the basis $\{\ket{g},\ket{e}\}$, the resonator state becomes a Schr\"odinger-cat-like superposition of orthogonally (opposite phase) squeezed states \begin{align}
    \ket{\Psi_\pm}&=\frac{1}{\mathcal{N_{\pm}}}(\ket{\zeta}\pm \ket{-\zeta}),
\end{align}
where $\mathcal{N}_\pm = \left[2(1\pm 1/\sqrt{\cosh(2r)})\right]^{1/2}$, and the sign depends on the measured qubit state. The Wigner function of the resonator state after measuring the qubit state in different bases is shown in Fig.~\ref{fig:WignerFunctions}. When the resonator is in a superposition of orthogonally squeezed states, its Wigner function dips to negative values in various regions of phase space, as shown in Fig.~\ref{fig:WignerFunctions}(b), thus making it a useful resource for non-Gaussian quantum computation \cite{ResourceWignerNeg}. The statistical and interference properties of general superpositions of squeezed states with different phases have been previously examined \cite{BCSanders_SqueezedSuperposition}. More recently, these states have been proposed as a resource for generating of heralded single photons \cite{azuma2024heralded}.
\begin{figure}[t]
        \includegraphics[scale=.8,left,trim={0 0 0 0}]{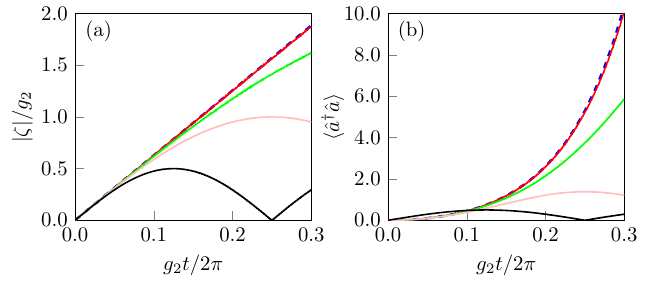}
        \caption{Dynamics of the squeezing parameter and photon number over time. The values of $\delta_2$ used are $\delta_2=2g_2$ (black), $\delta_2=g_2$ (pink), $\delta_2=0.5 g_2$ (green), $\delta_2=0.1 g_2$ (red) and $\delta_2=0$ (blue dashed). For a fixed value of $g_2$, the squeezing and, consequently, the photon number grow larger in time as $\delta_2$ goes to zero. {As in Fig.~\ref{fig:WignerFunctions}, we set $\Delta=0$ so that $\overline{g}_2=g_2$.} }
        \label{fig:ParameterDynamics}
\end{figure}

As mentioned above, when the qubit driving is two-photon-cross-resonant with the resonator ($\delta_2=0$), the squeezing parameter, $\zeta$, grows linearly in time. This leads to an exponential growth of the resonator photon number in time,
\begin{align}
    \bra{\pm \zeta(t)}\hat{a}^\dagger \hat{a}\ket{\pm\zeta(t)}=\sinh^2(\overline{g}_2t).
\end{align}
Figure~\ref{fig:ParameterDynamics} displays how the squeezing parameter and photon number change as a function of time for a fixed $\overline{g}_2$ and varying $\delta_2$. When $\delta_2 \ll \overline{g}_2$, $\zeta$ behaves similarly to the two-photon-cross-resonant case.

Henceforth, we refer to the procedure of applying Eq.~\eqref{eq:QCSTimeOp} as the QCS protocol. Interestingly, this protocol allows for the encoding of an arbitrary qubit state in a superposition of orthogonally squeezed states,  akin to how qubit states can be encoded using coherent states in bosonic cat codes \cite{CochraneCatCode,qcMAP,girvin2017schrodinger}.
{A peculiar feature of the superpositions of orthogonally squeezed states with opposite relative phases {$(\ket{\Psi_\pm})$} is that they {belong to} different Fock subspaces. This can be seen by writing these states in the Fock basis as \begin{align}
    \ket{\Psi_\pm}&=\frac{1}{\mathcal{N}_\pm} \sum_{n=0}^\infty \frac{\sqrt{(2n)!}}{2^n n!}(e^{i\phi}\tanh(r))^n ( (-1)^n \pm 1) \ket{2n}.
\end{align}
As shown in Ref.~\cite{BCSanders_SqueezedSuperposition}, depending on the relative phase between the two orthogonally squeezed states, the odd coefficients vanish for the plus sign and the state belongs to the even-two-photon multiples (four-photon) subspace spanned by $\{\ket{4n}\}$, where $n=0, 1, 2, ...$. Meanwhile, the even coefficients vanish for the minus sign, and the state belongs to the odd-two-photon multiples subspace spanned by $\{ \ket{4n+2}\}$. These states, $\ket{\Psi_+}$ and $\ket{\Psi_-}$, are orthogonal ($\braket{\Psi_\pm}{\Psi_\mp}=0$) and can be used to encode a logical qubit where we can take $\ket{0_{\text{L}}}=\ket{\Psi_+}$ and $\ket{1_{\text{L}}}=\ket{\Psi_-}$. The transition between the $\ket{0_{\text{L}}}$ and $\ket{1_{\text{L}}}$ subspaces can be implemented using two-photon jumps, i.e.~application of the operators $\hat{a}^{\dagger 2}$ and $\hat{a}^2$. Using a simple parity (non-destructive) measurement of the resonator, as done in the usual dispersive readout relying on $\sigZ\adag\aop$, one could infer when a two-photon jump has occured. Using the QCS protocol, we can generate one of the logical qubit states by measuring the qubit in Eq.~\eqref{eq:TimeEvSt} and depending on the measurement outcome we get $\ket{0_{\text{L}}}$ or $\ket{1_\text{L}}$. We now outline how to prepare an arbitrary logically-encoded qubit state. First, note that the Pauli-X logical states are
\begin{align}
    \ket{\pm_{\text{L}}}&=\frac{1}{\sqrt{2}}(\ket{0_{\text{L}}}\pm\ket{1_{\text{L}}})\nonumber\\ &= \frac{1}{\sqrt{2}\mathcal{N}_+}(\ket{\zeta} +\ket{-\zeta} ) \pm \frac{1}{\sqrt{2}\mathcal{N}_-}(\ket{\zeta} -\ket{-\zeta})\nonumber\\ &= \ket{\zeta}\left(\frac{1}{\sqrt{2}\mathcal{N}_+}\pm \frac{1}{\sqrt{2}\mathcal{N}_-}\right) \nonumber\\ &\,\,\,\,\,\,+\ket{-\zeta}\left(\frac{1}{\sqrt{2}\mathcal{N}_+}\mp\frac{1}{\sqrt{2}\mathcal{N}_-}\right).
\end{align}
In the limit of large squeezing, $\mathcal{N}_{\pm}\approx \sqrt{2}$ which means that $\ket{\pm_{\text{L}}}\simeq\ket{\pm\zeta}$. Thus, by preparing an arbitary qubit state $c_{+}\ket{+}+c_{-}\ket{-}$ and the resonator in vacuum, then performing the QCS protocol and measuring the qubit in the bare basis, we leave the resonator in the logically encoded state $\propto c_+\ket{+_{\text{L}}}\pm c_-\ket{-_{\text{L}}}$ with the relative phase depending the measurement outcome. Even when the squeezing is not large enough for this approximation, the finite sums and differences, $(\mathcal{N}_+ \pm \mathcal{N}_-)$, can be incorporated into the coefficients of the qubit state. From the definition of the logical states, we can construct logical one- and two-qubit gates. The form of these gates is somewhat strange since they are expressed in terms of squeezed states, e.g. \begin{align*}\hat{\sigma}_{x,\text{L}}&=\dyad{0_{\text{L}}}{1_{\text{L}}}+\dyad{1_{\text{L}}}{0_{\text{L}}}\nonumber\\ &\propto \dyad{\zeta}{\zeta}-\dyad{-\zeta}{-\zeta}+\dyad{\zeta}{-\zeta}-\dyad{-\zeta}{\zeta}.\end{align*} However, with universal control over the resonator Hilbert space, one can synthesize arbitrary unitaries and so this should not be a problem for state-of-the-art setups. The proposed encoding here serves as a two-photon generalization of the original bosonic damping cat code in Ref.~\cite{CochraneCatCode}. This {proposal} should pave the way for further research into the quantum error correction protocols associated with this code. }

{It is worth noting that the use of squeezing in augmenting existing quantum error correction codes is an active area of research and has been found to be useful in some cases, e.g. squeezed cat codes \cite{SqueezedCats}.}

\begin{figure}[t]
        \includegraphics[scale=.95,trim={1cm .8cm 0 0}]{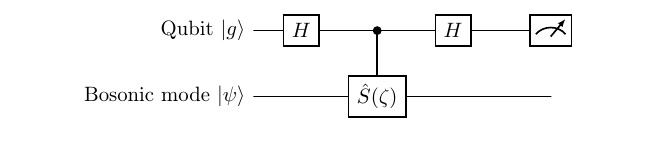}
        \caption{{Bosonic phase estimation using a controlled-squeeze gate which can be implemented with a single QCS operation (up to an unconditional squeezing). This circuit is the two-photon generalization of the controlled-displacement-based bosonic phase estimation protocol in Refs.~\cite{PhaseEstimationTerhal1} and~\cite{PhaseEstimationTerhal2}. The QCS operation can be decomposed into a global squeeze gate (not shown here) followed by a controlled-squeeze gate all sandwiched with a qubit Hadamard gate. If the QCS operation is performed in the bare basis ($\Delta\neq 0$ and $\Delta \gg \Omega$ ), the decomposition does not have the Hadamard gates and they must be added.}}
        \label{fig:PhaseEstimation}
\end{figure}
{
\subsubsection{Bosonic phase estimation using a controlled-squeeze gate}
A unitary operator, $\hat{U}$, has eigenvalues 
 of the form $e^{i\varphi_k}$ with respective eigenstates satisfying $\hat{U}\ket{\varphi_k}=e^{i\varphi_k}\ket{\varphi_k}$. {The basic idea of the phase estimation algorithm is to perform a unitary operator $\hat{U}$ conditioned on the state of an ancilla qubit, which enables the measurement of the eigenvalue, $e^{i\varphi_k}$ and the projection of any input state onto $\ket{\varphi_k}$. Phase estimation is typically done using two systems, an ancilla qubit and a target register of qubits. The controlled unitary $\hat{U}$ is implemented using standard quantum gates, and the measurement is performed on the ancilla qubit.} This can be generally done using any two quantum systems; the system types can be both continuous variable, both discrete variable, or a discrete-continous variable hybrid (in either permutation).}

 {A prototypical example of phase estimation can be seen in Fig.~\ref{fig:PhaseEstimation} where $\hat{S}(\zeta)$ is our specific target unitary of interest (as we discuss below), and generally it can be replaced with an arbitrary $\hat{U}$.  Note that there are different variants of the phase estimation circuit, e.g. some include a qubit rotation before the measurement to correct previous rounds of the protocol. For an overview of the different variants see the review and comparisons of various phase estimation protocols in Ref.~\cite{FastPhaseEstimation}. Phase estimation has many crucial applications such as quantum algorithms \cite{PhaseEstimLinearEq,lloyd2020quantumpolardecompositionalgorithm} including Shor's prime factorization algorithm \cite{ShorFactorization}, ground-state energy estimation \cite{GroundStatePhaseEstim}, and synchronizing clocks \cite{SyncPhaseEstim}.} 
 
 {Bosonic phase estimation refers to a setup where the target unitary is applied to a single or many bosonic modes. An important case of bosonic phase estimation is that of the displacement operator, $\hat{D}(\alpha)$. {The time-evolution operator of Eq.~\eqref{eq:GenTimeEvOp} for the case of $n=1$ is the qubit-conditional displacement (QCD) operation \begin{align}
     \hat{U}_{\text{QCD}}(\alpha)=\dyad{+}\hat{D}(\alpha) + \dyad{-}\hat{D}(-\alpha),
 \end{align} where the time dependence is kept implicit in $\alpha$.} This operation can be decomposed into a global displacement followed by a controlled-displacement which allows for the phase estimation of the displacement operator; 
 $$\hat{U}_{\text{QCD}}=H\hat{D}(\alpha)\widehat{CD}(-2\alpha)H,$$where $H$ is the qubit Hadanard gate and $$\widehat{CD}(\alpha)=\dyad{g}+\dyad{e}\hat{D}(\alpha)$$ is the controlled-displacement gate. The controlled-displacement phase estimation protocol has been studied in Ref.~\cite{PhaseEstimationTerhal1}.}

{Here, we generalize the displacement phase estimation protocol \cite{PhaseEstimationTerhal1} to the squeezing operator using a decomposition of the QCS operation into a composition of a global squeeze and controlled-squeeze gate. We write the QCS operation, the time-evolution operator in Eq.~\eqref{eq:QCSTimeOp}, while only keeping the $\zeta$ dependence and hiding the time dependence as implicit,}
{\begin{align}
    \hat{U}_{\text{QCS}}(\zeta)=\dyad{+}\hat{S}(\zeta) +\dyad{-}\hat{S}(-\zeta).
\end{align}
We define a controlled-squeeze gate that is controlled by the qubit as \begin{align}
    \widehat{CS}(\zeta)=\dyad{g} + \dyad{e}\hat{S}(\zeta).
\end{align}}
{Then, we can decompose the QCS unitary into a global squeeze operation followed by a controlled-squeeze operation all sandwiched by a qubit Hadamard gate:
\begin{align}
    \hat{U}_{\text{QCS}}(\zeta)=H\hat{S}(\zeta)\widehat{CS}(-2\zeta)H.
\end{align}}{With this decomposition, we can straightforwardly use QCS for bosonic phase estimation \footnote{Note that in the phase estimation circuit of Fig.~\ref{fig:PhaseEstimation}, we drop the unconditional global squeezing from the decomposition. Since it is unconditional, its effects can be accounted for at the end of the protocol as shown in Refs.~\cite{PhaseEstimationTerhal1,PhaseEstimationTerhal2}.}. In particular, with a repeated application of the circuit in Fig.~\ref{fig:PhaseEstimation}, the resonator state is projected onto an approximate eigenstate of the squeezing operator, $\hat{S}(\zeta)$, and the approximate eigenstate improves after each round  \cite{PhaseEstimationTerhal1,PhaseEstimationTerhal2}. If we, once again, consider a qubit interacting with a resonator simultaneously through a one- and two-photon interaction, one can then perform bosonic phase estimation concatenating controlled-displacement and controlled-squeeze allowing for the eigenvalue (phase) estimation of the concatenated operator, $\hat{D}(\alpha)\hat{S}(\zeta)$ (or $\hat{S}(\zeta)\hat{D}(\alpha)$) \cite{EigenSqueeze}. As mentioned above, phase estimation can be used for ground state energy (eigenvalue) estimation of a given Hamiltonian. With our controlled-squeeze phase estimation circuit and its concatenation with controlled-displacement, we can generally estimate the ground state energy of a bosonic Hamiltonian of the form $\hat{H}= \hbar (\xi_1 \adag + \xi_1^* \aop +\xi_2 \adagT + \xi_2^* \aop^2)$. {This can be {achieved when the} qubit interacts with the resonator simultaneously through one- and two-photon interactions, i.e.~with an interaction Hamiltonian of the form $$\hat{H}_I=\hbar g_1\sigX (\adag+\aop) + \hbar g_2\sigX(\adagT + \aop^2).$$ Then, the qubit can be tuned to be one-photon resonant when we implement the QCD operation where the two-photon interaction can be ignored. On the other hand, we can tune it to the two-photon resonance to implement the QCS operation. } }

{As a worthwhile side note, we {comment on the relationship between} bosonic phase estimation and quantum error correction. The study of Ref.~\cite{PhaseEstimationTerhal1} connects the displacement phase estimation protocol to the generation of Gottesman-Kitaev-Preskill (GKP) logical states \cite{GKPPaper2001}. The ideal GKP logical code states are unnormalizable states composed of an infinite superposition of displaced states quadrature ($\hat{x}$ or $\hat{p}$) eigenstates which are eigenstates of the displacement operators $\{\hat{D}(\alpha),\hat{D}(\beta)\}$ with $\alpha$ and $\beta$ together defining a GKP lattice, e.g. for a square GKP lattice, $\alpha=\sqrt{2\pi}$ and $\beta=i\sqrt{2\pi}$ which defines the desired commutation relation between $\hat{D}(\alpha)$ and $\hat{D}(\beta)$ \cite{GKPPaper2001}. Approximate forms of these states are found by replacing each displaced quadrature eigenstate with a displaced highly squeezed vacuum state and placing a Gaussian envelope over them - making the states of finite energy and, thus, physical. The displacement phase estimation protocol projects an input state onto the eigenstate of the displacement operator, and the repeated application of this phase estimation iteratively yields a better approximation of the $\hat{D}(\alpha)$ eigenstate which when $\alpha$ is properly selected, yields an approximate GKP state \cite{PhaseEstimationTerhal1,PhaseEstimationTerhal2}. The GKP code protects against small shift, i.e. displacement, errors or errors that can be decomposed into small shifts, {which include photon loss or dephasing}.  Applying the same insight into our proposed squeezing phase estimation protocol, the repeated application of this protocol yields an approximate eigenstate of $\hat{S}(\zeta)$. Using $SU(1,1)$ Lie-algebraic decomposition properties, one can show that the composition of squeezing operators in certain cases obeys $\hat{S}(\zeta_1)\hat{S}(\zeta_2)=\hat{S}(\zeta_3(\zeta_1,\zeta_2))\hat{R}(\theta(\zeta_1,\zeta_2)) $ with $\hat{R}(\theta)=\exp(i\theta\adag\aop)$ \cite{SU11Decomposition}. Essentially, two squeezes, under proper selection of squeezing parameters, $\zeta_1$ and $\zeta_2$, composed with each other yield a rotation followed by a squeeze. This identity can then be used to construct desired commutation relations, $[\hat{S}(\zeta_1),\hat{S}(\zeta_2)]$, for particular $\zeta_1$ and $\zeta_2$ values which would define a generalization of the GKP lattice to squeezing operators.  This Lie-algebraic identity can be used to formulate generalized desired commutation relations analogous to those of the displacement operators first formulated in \cite{GKPPaper2001}. This presents an opportunity to define an approximate bosonic error correction code that can correct small `squeezes', i.e.~squeezing errors, and other errors that can be decomposed into small squeezes. }

{\subsubsection{Expanding the generating set for universal qubit-resonator control}}
{In Ref.~\cite{FastUnivControlDisp}, the qubit-conditional displacement Hamiltonian (Eq.~\eqref{eq:EffHam} for $n=1)$ together with qubit rotations are shown to allow for (approximate) universal control of the joint qubit-resonator Hilbert space on short timescales, i.e. using a sequence of qubit-conditional displacements and qubit rotations one can (approximately) generate any arbitrary unitary 
 on the total Hilbert space. Defining the generalized position and momentum operators as $\hat{x}=(\adag +\aop)/\sqrt{2}$ and $\hat{p}=i(\adag -\aop)/\sqrt{2}$, respectively, the universal control can be though of as generating all possible operators of the form $\hat{\sigma}_j \hat{x}^k \hat{p}^l$ where $\hat{\sigma}\in\{\hat{\mathbb{I}},\sigX,\sigY,\sigZ\}$ and $j,k$ are non-negative integers.} 
 
 {For a generating set of Hamiltonians $\{\hat{H}_1,\hat{H}_2\}$, the short-timescale (short time steps $\delta t$ in the limit $\delta t \rightarrow 0$) universal control follows from the repeated application of following identities \cite{FastUnivControlDisp,Park2017_RepeatedIdentities}}{\begin{subequations}\label{eq:CommutatorIdentities}\begin{align}
     e^{-i\hat{H}_1 \tau}e^{-i\hat{H}_2 \tau}e^{i\hat{H}_1 \tau}e^{i\hat{H}_2 \tau}=e^{[\hat{H}_1,\hat{H}_2]\tau^2} + O(\tau^3),\label{eq:CommutatorIdentitiesA}\\\
     e^{i\hat{H}_1 \tau/2}e^{i\hat{H}_2 \tau/2}e^{i\hat{H}_2 \tau/2}e^{i\hat{H}_1 \tau/2}=e^{i(\hat{H}_1 +\hat{H}_2)\tau} + O(\tau^3),\label{eq:CommutatorIdentitiesB}\
 \end{align}\end{subequations}}
 {where $\tau=\delta t / \hbar$. With these identities, one can generate arbitrary superpositions of nested commutators from the generating set of Hamiltonians. In Ref.~\cite{FastUnivControlDisp}, the set of generators is that of conditional displacements and qubit rotations, $\mathcal{G}_1=\{\sigZ \hat{x}, \sigZ \hat{p},\sigX,\sigY,\sigZ  \}.$ This set is universal and can generate any operator of the form $\hat{\sigma}_j \hat{x}^k \hat{p}^l$.}
 
 {Here, we propose the inclusion of the generators of conditional squeezing, $\sigZ(\adagT+ \aop^2)$ and $i\sigZ(\adagT - \aop^2)$ \footnote{In general, we have access to the qubit-conditional squeezing generator $\hat{\sigma}_j(\eta \aop^2 - \eta^* \adagT )$ with $j=x,y,z$ and $\eta\in \mathbb{C}$. We choose to restrict our attention to a particular qubit operator $\sigZ$ and two orthogonal squeezing axes in phase space ($\adagT +\aop^2$ and $i(\adagT - \aop^2)$) to simplify the discussion and notation.}, which naturally {enable reaching} higher order polynomials in fewer steps and, thus, accumulating smaller errors due to the approximate nature of the identities in Eq.~\eqref{eq:CommutatorIdentities}.}
 
 {We can rewrite these generators using $\hat{x}$ and $\hat{p}$ as $\sigZ(\hat{x}^2 - \hat{p}^2)$ and $\sigZ(\hat{x}\hat{p} +\hat{p}\hat{x})=\sigZ\{\hat{x},\hat{p}\}$. We now show the steps required  to obtain the generators $\sigZ(\hat{x}^2 -\hat{p}^2)$ and $\sigZ\{\hat{x},\hat{p}\}$ using elements of the universal set $\mathcal{G}_1$. To realize the first generator, we need to obtain $\sigZ\hat{x}^2$ and $-\sigZ\hat{p}^2$ which we can then get the sum of using Eq.~\eqref{eq:CommutatorIdentitiesA}. Note that $\sigZ\hat{x}^2\propto [\sigX\hat{x},\sigY\hat{x}]$ and $\sigZ\hat{p}^2\propto [\sigX\hat{p},\sigY\hat{p}]$ which are themselves nested commutators since $\sigX\hat{x}\propto [\sigY,\sigZ\hat{x}]$ and $\sigX\hat{p}\propto [\sigY,\sigZ\hat{p}]$ (similarly for $\sigY\hat{x}$ and $\sigY\hat{p}$). Thus, first we must generate $\hat{\sigma}_j \hat{x}$ (and $\hat{\sigma}_j \hat{p}$) for $j=x,y$ which requires applying 2 qubit rotations and 2 conditional displacements for each $j$ to use Eq.~\eqref{eq:CommutatorIdentitiesA}, i.e. 4 operations to get each generator $\hat{\sigma}_j \hat{x}$ or $\hat{\sigma}_j \hat{p}$. Then, to obtain the generator $\hat{\sigma}_z \hat{x}^2$ ($\hat{\sigma}_z \hat{p}^2$), we must apply the unitary generated by $\hat{\sigma}_x \hat{x}$ twice and that generated by $\hat{\sigma}_y \hat{x}$ twice to use Eq.~\eqref{eq:CommutatorIdentitiesA} which in total requires 16 operations and similarly for $\sigZ\hat{p}^2$. Finally, to obtain the desired generator $\sigZ(\hat{x}^2 -\hat{p}^2)$, we need 64 operations since we must apply the 16 operations four times to use Eq.~\eqref{eq:CommutatorIdentitiesB}. A similar counting argument applies to obtaining the generator $\sigZ\{\hat{x},\hat{p}\}$.}
 
 {For any higher order target generator in the Lie algebra requiring the generators $\sigZ(\hat{x}^2 -\hat{p}^2)$ and $\sigZ\{\hat{x},\hat{p}\}$ as an intermediate step, it can be obtained using far less operations by having native access to these generators.  Additionally, reducing the circuit depth allows for more operations during the coherence lifetime of the device. The (redundant) generating set $\mathcal{G}_2=\{\sigZ \hat{x}, \sigZ \hat{p},\sigZ(\hat{x}^2 -\hat{p}^2),\sigZ\{\hat{x},\hat{p}\},\sigX,\sigY,\sigZ  \}$ helps synthesize target unitaries generated by higher order joint qubit-resonator operators more efficiently than the generating set $\mathcal{G}_1$, and it can be natively obtained through a qubit interacting with a resonator simultaneously through a one- and two-photon interaction (along with qubit rotations) as noted earlier.}
 \\

{\subsection{Higher-order interactions}}
{We can extend the above derivations and applications to higher order interactions with $n>2$.}

{
Effective qubit-conditional generalized $n$-photon squeezing operators as in Eq.~\eqref{eq:GenTimeEvOp} can be generally decomposed into a circuit involving a controlled-generalized-$n$-photon-squeezing gate. This controlled gate can be arranged in circuit as in Fig.~\ref{fig:PhaseEstimation} which can be used in single-shot or repeated bosonic phase estimation protocol of the generalized $n$-photon squeezing operators.}

{
The unitary synthesis advantage provided by the generators of conditional squeezing holds true for generalized squeezing as well. For the $n$-photon case, the two generators on orthogonal axes in phases space are $\sigZ(\adagn +\aop^n)$ and $i\sigZ(\adagn - \aop^n)$ which in terms of $\hat{x}$ and $\hat{p}$ can be written as $\sigZ((\hat{x}-i\hat{p})^n +(\hat{x}+i\hat{p})^n)/\sqrt{2^n}$ and $i\sigZ((\hat{x}-i\hat{p})^n -(\hat{x}+i\hat{p})^n)/\sqrt{2^n}$, respectively. Unsurprisingly, having native access to these generators provides a shortcut to synthesizing unitaries generated by higher order qubit-resonator operators.}

Finally, an unexplored research direction is the use of opposite phase generalized $n$-photon squeezed states for encoding a qubit as done here for $n=2$ and previously for $n=1$ in Ref.~\cite{CochraneCatCode}. One can investigate the Fock subspaces which these superposition states belong to and find out their orthogonality conditions. Similarly to the cases $n=1$ and $n=2$, the states with a `+' relative phase, $\propto(\hat{S}_n(\lambda) + \hat{S}_n(-\lambda))\ket{0}$, contain only even multiples of $n$ photons and hence belong to the subspace spanned by $\{\ket{2kn}\}$, where $k=0,1,2,...$, while the states $\propto(\hat{S}_n(\lambda) - \hat{S}_n(-\lambda))\ket{0}$ contain only odd multiples of $n$ photons and hence belong to the subspace spanned by $\{\ket{(2k+1)n}\}$. This property make these states viable candidates for a logical qubit encoding: owing to the orthogonality and good separation between the Fock states in these two subspaces, we can for example take the `+' relative phase state to be $\ket{0_{\text{L}}}$ and the `$-$' state to be $\ket{1_{\text{L}}}$. One difficulty with the theoretical treatment of these states is that we do not have explicit analytic expressions for the states in the Fock basis, and the power series expansion of the generalized squeezing operators $\hat{S}_n(\lambda)$ does not converge for $n>2$ \cite{SqueezingDivergence}. However, it has been shown that the operators' matrix elements can be numerically obtained for small squeezing parameters \cite{GeneralizedSqueezing,SqueezingStats}.  
\\

\section{Circuit QED implementation}\label{sec:cQED}

While the experiments proposed in Secs.~\ref{sec:TwoPhApp} and \ref{sec:QuantumSim} are implementation independent, we are interested in circuit QED as an implementation platform due to the range of coupling strengths it can achieve and its potential for scalability. There are two proposals in the literature for a circuit implementation of the two-photon Rabi model; a flux qubit inductively coupled to a superconducting quantum interference device (SQUID) \cite{ImplementationSC1,ImplementationSC2} and a split-Cooper-pair-box (charge qubit) inductively coupled to a transmission line \cite{ImplementationSC3}. In fact, the use of a (dc or rf) SQUID as a tunable coupler has long been known (see e.g. Refs.~\cite{SQUIDCoupler_00,SQUIDCoupler_01}), even though employing it to obtain nonperturbative nonlinear interactions is fairly recent \cite{CWilson2,JJEmbedded}.

\begin{figure}[t]
        \includegraphics[scale=.8,trim={0cm 0cm 0 0}]{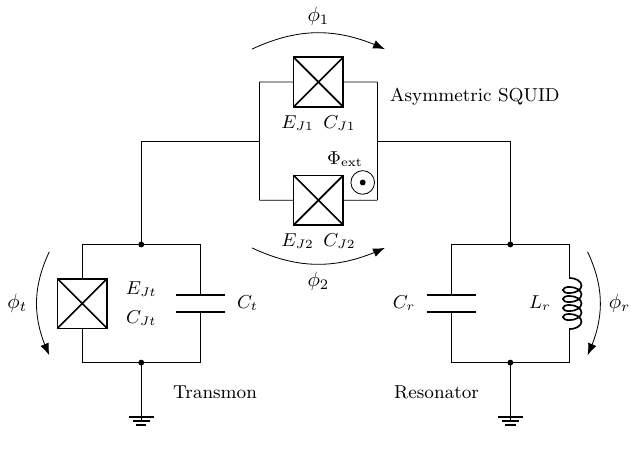}
        \caption{Transmon coupled to a resonator via an asymmetric SQUID. The transmon is characterized by a Josephson energy $E_{Jt}$ and charging energy $E_{Ct}=2e/(C_{Jt}+C_{t})$ depending on the junction and shunt capactiance; $E_{Jt}/E_{Ct}$ for this device is above 50 to operate in the transmon regime. The asymmetric SQUID has differing Josephson energies, $E_{J1}$ and $E_{J2}$, to exclusively allow for odd-order terms. The flux degrees of freedom are shown for the transmon, $\phi_t$, resonator, $\phi_r$, and the SQUID (coupler), $\phi_1$ and $\phi_2$. The coupler degrees of freedom are a function of the transmon, resonator and external flux threading the SQUID. }
        \label{fig:cQEDImplement}
\end{figure}
{\subsection{Two-photon operation mode}}
Here, we propose to employ the more widely used transmon qubit \cite{TransmonPaper} coupled to a lumped-element LC resonator via an asymmetric DC-SQUID threaded with an external flux, as shown in Fig.~\ref{fig:cQEDImplement}. The circuit Hamiltonian is (see App.~\ref{app:cQED} for a detailed derivation)
\begin{align}\label{eq:cQEDHam}
    \hat{H}=&\frac{\hat{q}_t^2}{2 \overline{C}_t}- E_{Jt}\cos(\frac{2\pi \hat{\phi}_t}{\phi_0}) + \frac{\hat{q}_r^2}{\overline{C}_r}+\frac{\hat{\phi}_r^2}{2 L_r} +\frac{1}{\overline{C}_c}\hat{q}_t \hat{q}_r\nonumber\\& - E_{c}\cos(\frac{2\pi ( \hat{\phi}_t-\hat{\phi}_r)}{\Phi_0})\nonumber\\  &-E_{s}\sin(\frac{2\pi ( \hat{\phi}_t-\hat{\phi}_r)}{\Phi_0}).
\end{align}
where $\Phi_{0}$ is the magnetic flux quantum, $\Phi_{\text{ext}}$ is the external flux threading the SQUID, and $\hat{\phi}_k$ and $\hat{q}_k$ are subsystem $k$'s flux and charge operators with $k=r$ referring to the resonator and $k=t$ referring to the transmon. The transmon is characterized by a Josephson energy $E_{Jt}$ and total (renormalized) capacitance $\overline{C}_t$, while the resonator is characterized by an inductance $L_r$ and (renormalized) capacitance $\overline{C}_r$. Finally, the SQUID coupler is characterized by asymmetric Josephson junctions with energies $E_{J1}$ and $E_{J2}$ dictating effective coupler energies $E_c =  E_{J1}\cos(2\pi\Phi_{\text{ext}}/{\Phi_0}) + E_{J2}$ and $ E_s =  E_{J1}\sin({2\pi\Phi_{\text{ext}}}/{\Phi_0})$. We assume that $E_{J1}\geq E_{J2}$. The asymmetry is necessary to exclusively generate odd-order qubit-resonator interactions, i.e., interactions of the form $\hat{\phi}_t^{j}\hat{\phi}_r^k$ where $j+k$ is an odd integer. We set $\Phi_{\text{ext}}=\Phi_0 \arccos(-E_{J2}/E_{J1})/2\pi$, thus, making $E_c=0$ such that the even order interactions are completely canceled. This results in a purely odd-order interaction. The two-photon JC interaction is classified under odd-order terms, generated by $\hat{\phi}_t\hat{\phi}_r^2$. When the qubit and resonator zero-point-fluctuation flux values are small, we may truncate the sine term, $\sin({2\pi ( \hat{\phi}_t-\hat{\phi}_r)}/{\Phi_0})$, at third order. Since the transmon is anharmonic and we assume the transition between the ground and first excited states to be the only resonant transition, the dynamics are confined to the lowest two energy eigenstates. Therefore, we also employ the \textit{two-level approximation} (TLA) such that the circuit QED Hamiltonian reads

\begin{table}[t]
\caption{\label{tab:table1}Circuit and Hamiltonian parameter estimates for operating a transmon coupled to a resonator via an asymmetric SQUID in the two-photon Jaynes-Cummings regime.}
\begin{ruledtabular}
\scalebox{1}{
\begin{tabular}{c c}
   Parameter&Two-photon mode\\
   \hline
   $\omega_q$&$2\pi\times$\SI{10}{\giga\hertz}\\
   $E_{Jt}/\hbar$&$2\pi\times$\SI{86.5}{\giga\hertz}\\
   $E_{Ct}/\hbar$&$2\pi\times$\SI{150}{\mega\hertz}\\
   $\omega_r$&$2\pi\times$\SI{5}{\giga \hertz}\\
   $C_r$&\SI{330}{\femto\farad}\\
   $E_{J1}/\hbar$&$2\pi\times$10.00-\SI{18.00}{\giga\hertz}\\
   $E_{J2}/\hbar$& $2\pi\times$9.94-\SI{17.50}{\giga\hertz}\\
   $g_2$ & $2\pi\times$25-\SI{50}{\mega\hertz}\\
   $\widetilde{g}_{e1}$ & $2\pi\times$1.08-\SI{2.16}{\giga\hertz} \\
   $\widetilde{g}_{e2}$& $2\pi\times$1.34-\SI{2.68}{\giga\hertz} \\
   $\widetilde{g}_{e3}$ & $2\pi\times$5-\SI{10}{\mega\hertz} \\
   $\widetilde{g}_{e4}$ & $2\pi\times$10-\SI{40}{\mega\hertz} \\
   $\widetilde{g}_{e5}$ & $2\pi\times$20-\SI{80}{\mega\hertz} \\
   $\widetilde{g}_{c}$ & $2\pi\times$30-\SI{50}{\mega\hertz} \\
   
\end{tabular}}
\end{ruledtabular}
\end{table}

\begin{align*}
    \hat{H}^{\text{TLA}}=&\frac{\hbar \omega_{q}}{2}\sigZ + \hbar \omega_r \adag \aop  - \hbar \widetilde{g}_{e4}(\adag + \aop)^3 \nonumber\\&- \hbar \widetilde{g}_{e5}\sigZ(\adag + \aop)+ \hbar g_{2} (\sigP + \sigM)(\adag+ \aop)^2 \nonumber\\ &- \hbar \widetilde{g}_c(\sigP -\sigM)(\adag - \aop),
    \nonumber\\ &- \hbar (\widetilde{g}_{e1}-\widetilde{g}_{e3})(\sigP + \sigM) \nonumber\\ &- \hbar (2\widetilde{g}_{e5}-\widetilde{g}_{e2})(\adag +\aop),
\end{align*}
where $\widetilde{g}_{e1},\,\widetilde{g}_{e2}\,\widetilde{g}_{e3},\,\widetilde{g}_{e4}$ and $\widetilde{g}_{e5}$ are spurious inductive couplings and $\widetilde{g}_{c}$ is a spurious capactive coupling. Here, $\omega_q=\sqrt{8E_{Ct} E_{Jt}}-E_{Ct}$ and $\omega_r=1/\sqrt{L_r\overline{C}_r}$. We now proceed to simplify this model to achieve the two-photon JC Hamiltonian. First, we neglect the linear offset terms, $\propto \sigP +\sigM$ and $\propto \adag+\aop$, as they can be tuned to zero \textit{in-situ} by applying a microwave field, and, thus, the tuned circuit QED Hamiltonian reads
\begin{align}\label{eq:TLAHamMain}
    \hat{H}^{\text{TLA}}=&\frac{\hbar \omega_{q}}{2}\sigZ + \hbar \omega_r \adag \aop  - \hbar \widetilde{g}_{e4}(\adag + \aop)^3 \nonumber\\&- \hbar \widetilde{g}_{e5}\sigZ(\adag + \aop)+ \hbar g_{2} (\sigP + \sigM)(\adag+ \aop)^2 \nonumber\\ &- \hbar \widetilde{g}_c(\sigP -\sigM)(\adag - \aop).
\end{align}
We now assume the two-photon JC conditions
\begin{align}
    2\omega_r=\omega_q \text{ and } g_2 \ll \omega_r.
\end{align}
With these conditions, only the two-photon JC terms are resonant, while all the other terms are off-resonant and can be neglected (see App.~\ref{app:cQED} for details). Then, the effective circuit QED Hamiltonian becomes
\begin{align}\label{eq:TwoPhJCMain}
    \hat{H}^{\text{TLA}}\simeq \frac{\hbar \omega_q}{2}\sigZ +\hbar\omega_r\adag \aop + \hbar g_{2}( \sigP \aop^2 + \sigM \adagT).
\end{align}
This final Hamiltonian shows that the proposed circuit has the necessary two-photon qubit-resonator interaction required to host the QCS protocol (and its subsequent applications) and to obtain an effective two-photon Rabi Hamiltonian at arbitrary coupling strengths.

\begin{figure}[t]
        \includegraphics[scale=.8,left,trim={0.17cm .8cm 0 0}]{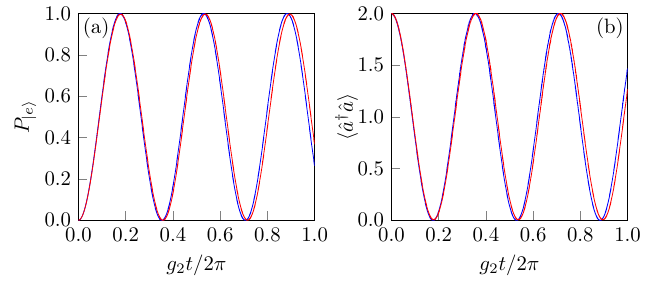}
        \caption{Two-photon Rabi oscillations exhibited in the dynamics of the qubit excited state probability and resonator photon number for an initial state $\ket{g}\ket{2}$. The blue lines are generated by the two-photon Jaynes-Cummings Hamiltonian in Eq.~\eqref{eq:TwoPhJCMain} and the red lines are generated by the two-level approximation circuit QED Hamiltonian in Eq.~\eqref{eq:TLAHamMain}. }
        \label{fig:TwoPhJC}
\end{figure}
In Table~\ref{tab:table1}, we provide realistic experimental parameters of the proposed circuit that can achieve the two-photon JC interaction. The details of the spurious couplings and their relations to the physical circuit parameters are derived in detail in App.~\ref{app:cQED}. We perform numerical simulations of the circuit QED model including spurious couplings using our estimated parameters to validate the two-photon JC interaction. Figure~\ref{fig:TwoPhJC} shows the dynamics generated using Eq.~\eqref{eq:TLAHamMain} contrasted to those generated using Eq.~\eqref{eq:TwoPhJCMain}. The probability of the excited state and the resonator photon number are shown as they evolve in time starting from an initial state $\ket{g}\ket{2}$. The circuit QED model exhibits two-photon Rabi oscillations in in excellent agreement with the ideal model.

The asymmetry of the SQUID we rely on here has also been used to implement multiphoton spontaneous parametric down conversion (SPDC) between bosonic modes \cite{CWilson1,CWilson2,JJEmbedded,CombinedConf,DynamicProtecCat}. The proposed device here is based on the same principles used for multiphoton SPDC with the difference being that we are coupling a bosonic mode to a transmon effectively truncated to its two lowest energy eigenstates. Our proposal can be used to reach the two-photon near-resonance strong coupling regime, i.e., $g_2\gg \kappa, \gamma_1,\gamma_\phi$ and $\omega_q\simeq 2\omega_r$, where $\kappa$ is the resonator photon loss rate, $\gamma_1$ is the qubit relaxation rate and $\gamma_\phi$ is the qubit dephasing rate. {Typical decoherence rates are all on the order of a few kHz while our coupling strength in the two-photon operation regime is on the order of tens of MHz, which is well into the strong coupling regime. } 

{\subsection{Higher-order interactions}}
The potential use for an asymmetric SQUID is not limited to the two-photon interactions. {Since the SQUID can host all orders of interactions it can, in principle, be used to obtain a specific order interaction with the qubit being tuned resonant with it, but the strength of the interaction terms rapidly diminish with higher orders. Even if we simply crank up the interaction strength through the SQUID's Josephson energies, each higher order interaction introduces its own unique problems such as qubit and resonator frequency renormalization, and resonant spurious terms that cannot be neglected with an RWA.}

{We briefly examine the issues arising from attempting to tune the system to realize a $n$-photon Jaynes-Cummings qubit-resonator interaction Hamiltonian, $\hat{H}_I=\hbar g_n(\sigP\aop^n + \sigM\adagn)$ for $n=3$ which illustrates the aforementioned issues. In the case of $n=3$, we rely on a fourth-order expansion of the SQUID interaction $E_{c}\cos({2\pi ( \hat{\phi}_t-\hat{\phi}_r)}/{\Phi_0})$  to obtain the term $\hat{\phi}_t\hat{\phi}_r^3$ which is proportional to $\sigP\adagTh +\sigM\aop^3$ under a three-photon resonance between the qubit and resonator. This term comes with other spurious terms; $\hat{\phi}_t^4$, $\hat{\phi}_r^4$, $\hat{\phi}_t^3\hat{\phi}_r$ and $\hat{\phi}_t^2\hat{\phi}_r^2$. Most of these spurious terms can be neglected since they are way off-resonance from the required three-photon resonance between the qubit and resonance. However, $\hat{\phi}_t^4$ and $\hat{\phi}_r^4$ introduce two issues: 1) they produce terms that renormalize the qubit and resonator frequencies which in turn alters the three-photon resonance condition, and 2) they produce resonant spurious Kerr-nonlinearities ($\propto \adagT\aop^2$) that need a careful treatment. These problems do not completely ruin the desired features, but they require more careful considerations and potentially slight modifications to the circuit. These problems similarly arise for orders higher than $n=3$. }
\\

\section{Summary and Conclusions}\label{sec:Conc}

To summarize, we presented a general theory on driving-enhanced $n$-photon qubit-resonator interactions. The multiphoton interactions are generated in the strong and qubit-detuned weak driving regimes with $n$-photon cross-resonance yielding the highest rate of generating photons in the resonator. After exploring the regimes of the driving-enhanced interactions, we explored the use of two drives to obtain an effective $n$-photon Rabi Hamiltonian with arbitrary coupling strength. When the first drive is the largest energy scale, the second drive plays the role of the qubit term in the effective Hamiltonian, while the detuning between the first drive and the resonator serves as the effective resonator frequency. Interestingly, the effective qubit-resonator $n$-photon coupling is given by the native coupling strength and is independent of the drive parameters.  

After developing the general framework for driving-enhance multiphoton qubit-resonator interactions, we focused on the case of $n=2$, where the theory yields qubit-conditional squeezing (QCS). Then, we described how the QCS protocol can be used in encoding a qubit state in the superposition of orthogonally squeezed states. {Furthermore, we outlined a controlled-squeeze bosonic phase estimation algorithm relying on QCS as well as its concatenation with controlled-displacement phase estimation. {Another potential application is} the expansion of the generating set for universal control of a qubit-resonator system using the generators of QCS which allows for more efficient unitary synthesis.}

From the implementation side, the generation of nonperturbative $n$-photon qubit-resonator interactions beyond $n=1$ is a challenging task. First, one major difficulty lies in obtaining a Hamiltonian where the $n^{\text{th}}$ order interaction can be isolated without the presence of spurious terms of comparable coupling strength. Second, the coupling strength in most systems significantly diminishes as the order of the interaction increases. Therefore, even if it is possible to obtain a Hamiltonian with the desired interaction, we require the strong coupling regime, i.e., $g_n \gg \kappa,\,\gamma_1,\,\gamma_\phi$. Without satisfying these conditions, the system will be dominated by losses, rendering the sought effects incoherent and suppressed by the system's losses. Here, we proposed an implementation that can achieve the necessary conditions for our theory in the case of $n=2$. The circuit uses a transmon qubit coupled to {an LC} resonator by means of an asymmetric SQUID. We provided realistic experimental parameters and validated the circuit QED model using numerical simulations exhibiting two-photon Rabi oscillations.

Throughout the paper, we exclusively discussed unitary evolution. In App.~\ref{app:decoherence}, we perform extensive open system numerical simulations for worse-than-average decoherence qubit and resonator parameters and corroborate the analytical results; the predictions are robust against qubit energy relaxation and dephasing and resonator photon loss. We find the fidelity of the states generated to be largely unaffected at the timescales of consideration used in the paper.

This work paves the way for a new set of nonperturbative multiphoton qubit-resonator effects that can be leveraged for applications in various quantum applications for information processing --- as presented here, sensing, and communication.

\begin{acknowledgments}
We thank L. Dellantonio and E. Peters for their feedback on an earlier version of this work. M.A. was supported by the Institute for Quantum Computing (IQC) through Transformative Quantum Technologies (TQT). M.A., X.X. and M.M. acknowledge funding
from the Canada First Research Excellence Fund (CFREF) and the support of the Natural Sciences and Engineering Research Council of Canada (NSERC) (Application
No. RGPIN-2019-04022). S.A. was supported by Japan's Ministry of Education, Culture, Sports, Science and Technology's Quantum Leap Flagship Program Grant No. JPMXS0120319794.
\\

{After completion of this manuscript, we became aware
of three manuscripts that discuss related proposals and
applications: Ref.~\cite{HybridOscillatorQubit} is a comprehensive review of hybrid qubit-oscillator architectures and their protocols with a segment dedicated to controlled-squeezing, Ref.~\cite{CtrlSqueeze} proposes a controlled-squeeze operation similar to our proposal but instead relying on the dispersive regime and a two-photon drive, and Ref.~\cite{SanerNonclassicalHOStates} reports the experimental realization of qubit-conditional squeezing, triqsqueezing and quadsqueezing in a trapped ion system.}
\end{acknowledgments}

\appendix
    \section{Decoherence}\label{app:decoherence}
     \begin{figure*}[t]
\includegraphics[trim={1cm 0 0 0},scale=1.1]{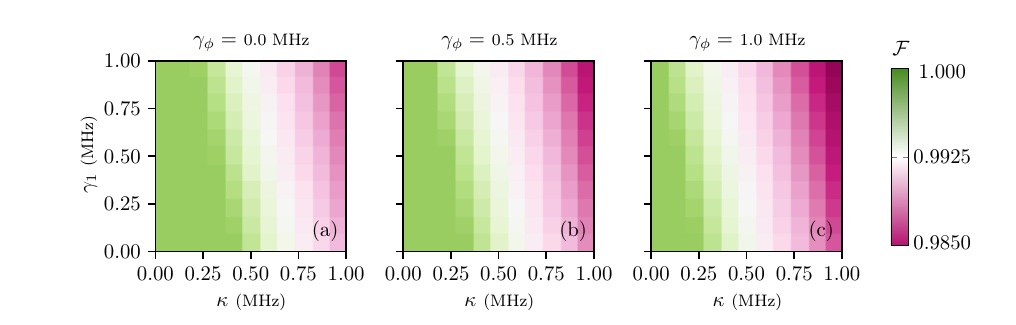}
\caption{\label{fig:Fidelity} {Fidelity of prepared state with varying decoherence rates. The Hamiltonian parameters used for the simulations are $\Omega=2\pi\times \SI{0.5}{\giga\hertz}$, $g_2=2\pi\times \SI{20}{\mega\hertz}$, and $\Delta=\delta_2=0$ (identical to those in Fig.~1 in the main text with the decoherence parameters varied) {with a joint initial state $\ket{g}\ket{0}$}. (a)-(c) The fidelity between a state prepared via time-evolution using Eq.\eqref{eq:QubitMasterEqn} and a reference state evolved with Eq.\eqref{eq:SysHam} is plotted for different qubit and resonator decoherence rates, with a time-evolution period of $g_2t/2\pi=0.3$. The resonator photon loss rate, $\kappa$, is the most detrimental parameter to the state fidelity. The qubit relaxation rate, $\gamma_1$, also diminishes the fidelity but the qubit dephasing, $\gamma_\phi$, is practically negligible as the three plots are nearly identical.}} 
\end{figure*}
    We only considered unitary time evolution in the main text. The qubit-resonator system are not completely isolated from the environment. Here, we take into account qubit energy relaxation, qubit dephasing and resonator photon loss. Since we are operating in the strong coupling regime, the qubit and resonator are not strongly hybridized and, thus, we can assume they interact with separate baths at zero temperature. For these conditions, it suffices to model the open system using a Lindblad master equation that reads  \cite{Breuer_Petruccione_2010,Carmichael_1993}
\begin{align}\label{eq:QubitMasterEqn}
    \frac{d}{dt}\hat{\rho}=-\frac{i}{\hbar}[\hat{H},\hat{\rho}] + \gamma_1 \mathcal{D}(\hat{\sigma}_{-})\hat{\rho} +  \frac{\gamma_{\phi}}{2} \mathcal{D}(\hat{\sigma}_z)\hat{\rho} + \kappa \mathcal{D}(\hat{a})\hat{\rho},
\end{align}
where $\hat{\rho}$ is the full system density matrix, $\mathcal{D}(\hat{O})\hat{\rho}=\hat{O}\hat{\rho}\hat{O}^\dagger - \{\hat{O}^\dagger \hat{O},\hat{\rho}\}/2$ is the dissipator for a given operator $\hat{O}$, $\gamma_1$ and $\gamma_\phi$ are the qubit energy relaxation and dephasing rates, and $\kappa$ is the resonator photon loss rate. In what follows we use the Python library QuTiP \cite{QuTiP}

We seek to evaluate the contribution of different decoherence parameters on the prepared state, $\hat{\rho}_{\text{prep}}$. For that purpose, we define the fidelity as $$\mathcal{F}=\left(\Tr(\sqrt{\sqrt{\hat{\rho}_{\text{prep}}}\hat{\rho}_{\text{ideal}}\sqrt{\hat{\rho}_{\text{prep}}}})\right)^2,$$
where we use an ideal reference state $\hat{\rho}_{\text{ideal}}$ obtained using  Eq.~\eqref{eq:SysHam} and $\hat{\rho}_{\text{prep}}$ is arrived at using the master equation, Eq.~\eqref{eq:QubitMasterEqn}. Figure~\ref{fig:Fidelity} demonstrates the results of numerical simulations in which both the reference and prepared states experienced time evolution over a normalized time of $g_2t/2\pi=0.3$. The same figure also highlights that the photon loss rate in the resonator is the most significant factor affecting the fidelity of the state. Although qubit relaxation contributes to a reduction in fidelity, the influence of qubit dephasing is almost inconsequential within this timescale. It is important to note that the maximum decoherence rates used in our simulations, set at $\SI{1}{\mega\hertz}$, are substantially higher than those typically found in present circuit QED setups, and even more so in state-of-the-art devices.

Note that the time-evolution period may seem short ($g_2 t/2\pi=0.3$), but by the analytic estimate of Eq.~(7) in the main text, the resonator will contain much more than 100 photons in a duration less than $g_2 t/2\pi=0.5$. This is also apparent by the form of the squeezing parameter, $\zeta=ig_2t$ which grows linearly in time. Due to this fact, it is very hard to simulate long-time dynamics using traditional software packages such as QuTiP (employed here). For these simulations, we truncated the resonator Hilbert space to 150 photons. A more rigorous numerical study is needed for the study of the long-time dynamics and eventual decay of the photon number, but for the purpose of ensuring the robustness of state preparation against decoherence, the simulations here suffice to corroborate the analytical predictions.

\section{Derivation of Circuit QED Implementation\label{app:cQED}}
In this section, we derive and quantize the system Hamiltonian for the circuit implementation proposed in the main text. We then proceed to apply the two-level approximation to the transmon along with the relevant RWA to obtain the two-photon Jaynes-Cummings Hamiltonian.

\subsection{Circuit Hamiltonian}

We first begin by stating the total system (transmon, resonator and coupler) Lagrangian for the circuit shown in Fig.~\ref{fig:cQEDImplement}. We use the system constraints to eliminate the coupler degree of freedom and express it in terms of the transmon and resonator degrees of freedom. Then, we obtain the classical Hamiltonian by means of a Legendre transformation. Then, we promote the conjugate variables to quantum operators, arriving at a quantum-mechanical description of the circuit.

The total system Lagrangian is \cite{CWilson2}
\begin{align}
    \mathcal{L}_{\text{total}}=\mathcal{L}_{\text{transmon}}+\mathcal{L}_{\text{resonator}}+\mathcal{L}_{\text{coupler}},
\end{align}
where
\begin{align}
    \mathcal{L}_{\text{transmon}}=\frac{1}{2}C_{\Sigma}\dot{\phi}_t^2 + E_{Jt}\cos(\frac{2\pi \phi_t}{\Phi_0}),
\end{align}
\begin{align}
    \mathcal{L}_{\text{resonator}}=\frac{1}{2}C_{r}\dot{\phi}_r^2 - \frac{1}{2L_r}\phi_r^2,
\end{align}
and
\begin{align}
    \mathcal{L}_{\text{coupler}}=&\frac{1}{2}C_{J1}\dot{\phi}_1^2 + \frac{1}{2}C_{J2}\dot{\phi}_2^2 \nonumber \\ & + E_{J1}\cos(\frac{2\pi \phi_1}{\Phi_0}) + E_{J2}\cos(\frac{2\pi \phi_2}{\Phi_0}). \label{eq:CouplerLagrangianOriginal}
\end{align}
Here, $\phi_j$ is the flux variable and $\dot{\phi}_j$ is its time derivative for the $j$ subsystem with $t$ denoting the transmon, $r$ denoting the resonator and 1 and 2 denoting the two junctions of the DC-SQUID coupler. The transmon is characterized by Josephson energy $E_{Jt}$ and total (junction and shunting capacitance) $C_{\Sigma}=C_{Jt}+C_{t}$. The resonator is characterized by the inductance $L_r$ and capacitance $C_r$. Lastly, the SQUID is characterized by its junction capacitances $C_{J1}$ and $C_{J2}$ and Josephson energies $E_{J1}$ and $E_{J2}$.

We now derive relations between the different circuit variables and use these relations as constraints to eliminate redundant variables. Specifically, we would like to eliminate the coupler and instead obtain a Lagrangian written in terms of the transmon and resonator variables only. For this purpose, we need to examine the circuit in Fig.~\ref{fig:cQEDImplement}. Assuming a constant external flux, the time-derivatives of the flux variables are related through Kirchoff's voltage law (KVL) by:
\begin{align}
    \dot{\phi}_1-\dot{\phi}_2=0. \label{eq:ConstraintSmallLoopDerivative}
\end{align}
and
\begin{align}
    \dot{\phi}_t-\dot{\phi}_2-\dot{\phi}_r=0 \label{eq:ConstraintLargeLoopDerivative}
\end{align}
Integrating these KVL constraints yields the flux relations
\begin{align}
    \phi_1-\phi_2=\widetilde{\Phi} \label{eq:ConstraintSmallLoop}
\end{align}
and
\begin{align}
    {\phi}_t-{\phi}_2-{\phi}_r = \widetilde{\widetilde{\Phi}}, \label{eq:ConstraintLargeLoop}
\end{align}
where $\tilde{\Phi}$ and $\widetilde{\widetilde{\Phi}}$ are constants of integration that are determined based on the fluxes in the different loops in the circuit. As mentioned above, a flux $\Phi_{\text{ext}}$ is applied to the SQUID loop. We assume that there is no external flux applied to the loop that goes from the ground through the transmon's junction, the coupler's bottom junction, the resonator's inductance and back to the ground. The constants $\tilde{\Phi}$ and $\widetilde{\widetilde{\Phi}}$ are then given by $\widetilde{\Phi}= - \Phi_{\text{ext}} + k_1 \Phi_0$ and $\widetilde{\widetilde{\Phi}}=k_2 \Phi_0$ for some integers $k_1$ and $k_2$. 
We now use Eqs.~(\ref{eq:ConstraintSmallLoopDerivative}) and (\ref{eq:ConstraintSmallLoop}) to eliminate $\phi_1$ from Eq.~(\ref{eq:CouplerLagrangianOriginal}) and obtain
\begin{align}
    \mathcal{L}_{\text{coupler}}=&\frac{1}{2}(C_{J1} + C_{J2})\dot{\phi}_2^2 + E_{J1}\cos(\frac{2\pi (\phi_2 -\Phi_{\text{ext}})}{\Phi_0}) \nonumber \\ & + E_{J2}\cos(\frac{2\pi \phi_2}{\Phi_0}) \nonumber \\
    =&\frac{1}{2}C_c\dot{\phi}_2^2 + E_c\cos(\frac{2\pi \phi_2}{\Phi_0}) + E_s\sin(\frac{2\pi \phi_2}{\Phi_0}), \label{eq:CouplerLagrangianIntermediate}
\end{align}
where
\begin{align}
    C_c = & C_{J1} + C_{J2}, \nonumber \\
    E_c = & E_{J1}\cos(\frac{2\pi\Phi_{\text{ext}}}{\Phi_0}) + E_{J2}, \nonumber \\
    E_s = &  E_{J1}\sin(\frac{2\pi\Phi_{\text{ext}}}{\Phi_0}).
\end{align}
Next we use Eqs.~(\ref{eq:ConstraintLargeLoopDerivative}) and (\ref{eq:ConstraintLargeLoop}) to eliminate $\phi_2$ from Eq.~(\ref{eq:CouplerLagrangianIntermediate}) and obtain
\begin{align}
    \mathcal{L}_{\text{coupler}}=&\frac{1}{2}C_c(\dot{\phi}_t-\dot{\phi}_r)^2 + E_c\cos(\frac{2\pi (\phi_t-\phi_r)}{\Phi_0}) \nonumber \\ & + E_s\sin(\frac{2\pi (\phi_t-\phi_r)}{\Phi_0}).
\end{align}

The total Lagrangian can then be written as
\begin{align}
    \mathcal{L}=\frac{1}{2}\dot{\vec{\phi}}^{\,T} \boldsymbol{C} \dot{\vec{\phi}} - U(\vec{\phi})
\end{align}
where
\begin{align}
    \vec{\phi}=\begin{pmatrix}
        \phi_t \\ \phi_r
    \end{pmatrix},
\end{align}
\begin{align}
\boldsymbol{C}=\begin{pmatrix}
                C_t+C_{Jt}+C_{J1}+C_{J2} & -(C_{J1}+C_{J2}) \\
                -(C_{J1}+C_{J2}) & C_r+C_{J1}+C_{J2}
\end{pmatrix},    
\end{align}
and
\begin{align}
    U(\vec{\phi})=& -E_{c}\cos(\frac{2\pi (\phi_t-\phi_r)}{\Phi_0})\nonumber\\ & + E_{s}\sin(\frac{2\pi ( \phi_t-\phi_r)}{\Phi_0})\nonumber\\ &-E_{Jt}\cos(\frac{2\pi \phi_t}{\Phi_0}) + \frac{1}{2L_r}\phi_r^2.
\end{align}
Then, the Hamiltonian can be derived via the Legendre transform as
\begin{align}
    H=\frac{1}{2}\vec{q}^{\,T} (\boldsymbol{C}^{-1})\vec{q}+ U(\vec{\phi}),
\end{align}
where
\begin{align}
    \vec{q}=\begin{pmatrix}
        q_t \\ q_r
    \end{pmatrix},
\end{align}
with $q_k=\partial \mathcal{L}/\partial \dot{\phi}_k$ ($k=q,r$) being the charges stored in the qubit and resonator. We label $(\boldsymbol{C}^{-1})_{11}\equiv \overline{C}_t^{-1}$, $(\boldsymbol{C}^{-1})_{22}\equiv \overline{C}_r^{-1}$, and $(\boldsymbol{C}^{-1})_{12}=(\boldsymbol{C}^{-1})_{21}\equiv \overline{C}_{c}^{-1}$. Thus, the Hamiltonian can be written as
\begin{align}
    H=&\frac{q_t^2}{2 \overline{C}_t}- E_{Jt}\cos(\frac{2\pi \phi_t}{\phi_0}) + \frac{q_r^2}{\overline{C}_r}+\frac{\phi_r^2}{2 L_r} +\frac{1}{\overline{C}_c}q_t q_r \nonumber\\ &- E_{c}\cos(\frac{2\pi ( \phi_t-\phi_r)}{\Phi_0})\nonumber\\  &+E_{s}\sin(\frac{2\pi ( \phi_t-\phi_r)}{\Phi_0}).
\end{align} 
Finally, we promote the classical Poisson brackets to commutator brackets via the rule $$\{\phi_t,q_t\}{=1}\mapsto [\hat{\phi}_t,\hat{q}_t]=i\hbar,$$
$$\{\phi_r,q_r\}{=1}\mapsto [\hat{\phi}_r,\hat{q}_r]=i\hbar,$$
where $\hat{q}_k,\,\hat{\phi}_k$ are now quantized operators. By reeplacing the variables with operators, we obtain the quantum circuit Hamiltonian in Eq.~\eqref{eq:cQEDHam}.

For the derivations to follow, we isolate the inductive and capacitive SQUID interactions from the total Hamiltonian. The inductive SQUID interaction Hamiltonian we refer to is
\begin{align}
    \hat{H}_{I,\text{SQ}}^{\text{ind}}=&- E_{c}\cos(\frac{2\pi ( \hat{\phi}_t-\hat{\phi}_r)}{\Phi_0})\nonumber\\  &-E_{s}\sin(\frac{2\pi ( \hat{\phi}_t-\hat{\phi}_r)}{\Phi_0}).
\end{align}
While, the capacitive SQUID interaction Hamiltonian is
\begin{align}
    \hat{H}_{I,\text{SQ}}^{\text{cap}}=\frac{1}{\overline{C}_c}\hat{q}_t \hat{q}_r.
\end{align}
Finally, we rewrite the qubit and resonator flux and charge operators using the bosonic creation and annihilation operators,
\begin{subequations}
\begin{align}
    \hat{q}_{r}=iq_{\text{zpf},r}(\adag -\aop),
\end{align}
\begin{align}
    \hat{\phi}_{r}=\phi_{\text{zpf},r}(\adag +\aop),
\end{align}
\begin{align}
    \hat{q}_{t}=iq_{\text{zpf},t}(\bdag -\bop),
\end{align}
and
\begin{align}
    \hat{\phi}_{t}=\phi_{\text{zpf},t}(\bdag +\bop),
\end{align}
\end{subequations}
where $\phi_{\text{zpf},t}$ and $q_{\text{zpf},t}$ ($\phi_{\text{zpf},r}$ and $q_{\text{zpf},r}$) are the qubit (resonator) zero-point-fluctuation flux and charge values, respectively. The commutation relations for the qubit and resonator creation and annihilation operators are $[\bop,\bdag]=\mathbb{I}$ and $[\aop,\adag]=\mathbb{I}$, respectively.
\\

\subsection{Two-photon Jaynes-Cummings Hamiltonian}

For small zero-point-fluctuation flux values, we may Taylor-expand the sine and cosine into the first few polynomial terms. We are interested in the odd order terms to obtain an effective two-photon Jaynes-Cummings Hamiltonian. For this purpose, we now focus on the odd order terms by setting $\Phi_{\text{ext}}=\Phi_0 \arccos(-E_{J2}/E_{J1})/2\pi$, thus, making $E_c=0$.
\begin{widetext}
\begin{align}
    \hat{H}_{I,\text{SQ}}^{\text{ind}}&=-E_{s}\left[\frac{2\pi(\hat{\phi}_t-\hat{\phi}_r)}{\Phi_0} -\frac{1}{3!}\frac{8\pi^3 (\hat{\phi}_t-\hat{\phi}_r)^3}{\Phi_0^3}\right]\nonumber\\ &= -E_{s}\Bigg[ \frac{2\pi}{\Phi_0}(\phi_{\text{zpf},t}(\bdag+\bop ) -\phi_{\text{zpf},r}(\adag +\aop) )-\frac{1}{3!}\frac{8\pi^3}{\Phi_0^3}\big(\phi_{\text{zpf},t}^{3} (\bdag+\bop)^3-\phi_{\text{zpf},r}^{3} (\adag +\aop)^3 \nonumber\\&\,\,\,\,\,-3\phi_{\text{zpf},t}^2\phi_{\text{zpf},r}(\bdag+\bop)^2(\adag + \aop) + 3\phi_{\text{zpf},t}\phi_{\text{zpf},r}^2(\bdag +\bop)(\adag +\aop)^2 \big)   \Bigg]
\end{align}
\end{widetext}
We now rearrange terms using the commutation relations, $[\bop,\bdag]=\mathbb{I}$ and $[\aop,\adag]=\mathbb{I}$, and we use the \textit{two-level approximation} where usually $\bop\mapsto\sigM\,\,(\bdag\mapsto\sigP)$ and $\bdag\bop\mapsto\sigZ$, and for the higher-order terms we truncate to the two-dimensional subspace. In this case, $\bdag+\bop\simeq \sigP+\sigM$, $(\bdag+\bop)^2\simeq \sigZ +2\hat{\mathbb{I}}$, $(\bdag+\bop)^3\simeq3(\sigP+\sigM)$ and $(\bdag+\bop)^4\simeq(9\sigZ+6\hat{\mathbb{I}})$. Thus, we get that
\begin{widetext}
\begin{align}
    \hat{H}_{\text{SQ},I}^{\text{ind,TLA}}&\simeq -E_s\Bigg[\frac{2\pi}{\Phi_0}(\phi_{\text{zpf},t}(\sigP+\sigM ))- \frac{2\pi}{\Phi_0}(\phi_{\text{zpf},r}(\adag+\aop ))-\frac{1}{3!}\frac{8\pi^3}{\Phi_0^3}\big(\phi_{\text{zpf},t}^3 3(\sigP +\sigM) -\phi_{\text{zpf},r}^3 (\adag +\aop)^3 \nonumber\\ &\,\,\,\,\,\,\,\,\,\,\,\,\,\,\,\,\,\,\,\,\,\,\,\,-3\phi_{\text{zpf},t}^2\phi_{\text{zpf},r}(\sigZ +2\hat{\mathbb{I}})(\adag +\aop) + 3\phi_{\text{zpf},t}\phi_{\text{zpf},r}^2(\sigP +\sigM)(\adag+\aop)^2\big)\Bigg]\nonumber\\ &= -\hbar \widetilde{g}_{e1}(\sigP + \sigM) + \hbar \widetilde{g}_{e2}(\adag +\aop)+ \hbar \widetilde{g}_{e3}(\sigP + \sigM) - \hbar \widetilde{g}_{e4}(\adag + \aop)^3 \nonumber\\&\,\,\,\,\, - \hbar \widetilde{g}_{e5}(\sigZ+2\hat{\mathbb{I}})(\adag + \aop) + \hbar g_{2} (\sigP + \sigM)(\adag+ \aop)^2,
\end{align}
\end{widetext}
where $\hbar \widetilde{g}_{e1} =E_{s}\eta_t$, $\hbar \widetilde{g}_{e2} =E_{s}\eta_r$, $\hbar \widetilde{g}_{e3} =3E_{s}\eta_t^3 /3!$, $\hbar \widetilde{g}_{e4} =E_{s}\eta_r^3/3!$, $\hbar \widetilde{g}_{e5}=3E_{s}\eta_t^2 \eta_r/3!$ and $\hbar g_{2}=3E_{s}\eta_t\eta_r^2/3!$. Here, $\eta_{t/r}=2\pi \phi_{\text{zpf},t/r}/\Phi_0$ is the ratio between the zero-point-fluctuation flux of the qubit (resonator) and the flux quantum. For simplicity, we drop the linear offset qubit and resonator terms, $\propto \sigP \pm \sigM$ and $\propto \adag \pm \aop$, since we can cancel them with a displacement that can be tuned in situ during the experiment.

Next, we turn our attention to the capacitive interaction term with the goal of obtaining the final two-level approximation form,
\begin{align}
    \hat{H}_{\text{SQ},I}^{\text{cap}}&=\frac{1}{\overline{C}_c}\hat{q}_t \hat{q}_r\simeq-\hbar \widetilde{g}_{c}(\sigP-\sigM)(\adag - \aop),
\end{align}
where $\hbar \widetilde{g}_{c}={q_{\text{zpf},t}q_{\text{zpf},r}}/{\overline{C}_c}$.
We now collect the bare and interaction terms to write down the full Hamiltonian in the two-level approximation.
\begin{align}\label{eq:TLAHam}
    \hat{H}^{\text{TLA}}=&\frac{\hbar \omega_{q}}{2}\sigZ + \hbar \omega_r \adag \aop  - \hbar \widetilde{g}_{e4}(\adag + \aop)^3 \nonumber\\&- \hbar \widetilde{g}_{e5}\sigZ(\adag + \aop)+ \hbar g_{2} (\sigP + \sigM)(\adag+ \aop)^2 \nonumber\\ &- \hbar \widetilde{g}_c(\sigP -\sigM)(\adag - \aop),
\end{align}
where $\omega_q=\sqrt{8E_{Ct} E_{Jt}}-E_{Ct}$ and $\omega_r=1/\sqrt{L_r\overline{C}_r}$. 
We now assume near two-photon resonance between the qubit and resonator, $\omega_q \simeq 2 \omega_r$. We can then transform to the usual rotating frame via $\hat{U}_r=\exp[-it(\omega_q \sigZ/2 + \omega_r \adag\aop)]$. In this frame the operators oscillate as
\begin{align*}
    &\sigM\mapsto \sigM e^{-i\omega_q t},\\
    & \aop\mapsto \aop e^{-i\omega_r t},
\end{align*}
which leads to the two-photon Jaynes-Cummings terms, $\sigP\adagT$ and $\sigM\aop^2$, being the only slow-rotating terms while everything else is fast-rotating. In particular, we require that
\begin{subequations}
\begin{align}
    \widetilde{g}_{e4},\widetilde{g}_{e5}\ll \omega_r,
\end{align}
\begin{align}
    \widetilde{g}_{c}\ll  |\omega_q-\omega_r|,\, \omega_q+\omega_r,
\end{align}
and
\begin{align}
    {g}_{2}\ll  \omega_q+2\omega_r.
\end{align}
\end{subequations}
Finally, imposing these conditions and dropping their associated terms, we arrive at the two-photon Jaynes-Cummings Hamiltonian
\begin{align}\label{eq:TwoPhJC}
    \hat{H}^{\text{TLA}}\simeq\hat{H}_{\text{2-JC}}=\frac{\hbar \omega_q}{2}\sigZ +\hbar\omega_r\adag \aop + \hbar g_{2}(\sigP \adagT + \sigM \aop^2).
\end{align}
This is the system Hamiltonian needed in Eq.~\eqref{eq:SimplifiedRotFrameHam} in the main text for the case of $n=2$.

\bibliography{paper}

\end{document}